%% file: MC_arXiv_MAIN.tex
\documentclass[10pt,journal,letterpaper]{IEEEtran}

\usepackage{cite}
\usepackage{amsmath,amssymb,amsfonts}
\usepackage{graphicx}
\usepackage{textcomp}
\usepackage{xcolor}
\usepackage{bm}
\usepackage{float}
\usepackage{algorithm}
\usepackage{algpseudocode}
\usepackage{wrapfig}
\usepackage{balance}
\usepackage{comment}
\usepackage{enumitem,kantlipsum}

\def\endthebibliography{%
	\def\@noitemerr{\@latex@warning{Empty `thebibliography' environment}}%
	\endlist
}
\def\BibTeX{{\rm B\kern-.05em{\sc i\kern-.025em b}\kern-.08em
		T\kern-.1667em\lower.7ex\hbox{E}\kern-.125emX}}

\usepackage{titling}
\renewcommand\maketitlehooka{
  \setlength\parindent{0pt}
  \begin{minipage}{\textwidth}
    \center
    This work has been submitted to the IEEE for possible publication.\\
    Copyright may be transferred without notice, after which this version may no longer be accessible.
  \end{minipage}
  \vspace{0.2in}
}
\date{}

\begin{document}

\title{Modeling and Analysis of Coexistence Between MLO NSTR-based Wi-Fi 7 and Legacy Wi-Fi}

\author{Suhwan Jung, Seokwoo Choi, Youngkeun Yoon, Ho-kyung Son, Hyoil Kim, {\it Senior Member, IEEE} 
\thanks{S. Jung, S. Choi and H. Kim are with Dept. of Electrical Engineering, Ulsan National Institute of Science and Technology (UNIST), Ulsan 44919, Republic of Korea (e-mail: \{fydnrltlzl,csw2055,hkim\}@unist.ac.kr).}
\thanks{Y. Yoon and H.-K. Son are with the Radio Research Division, Electronics and Telecommunications Research Institute (ETRI), Daejeon 34129, Republic of Korea (email: \{ykyoon,hgson\}@etri.re.kr).}
\thanks{Hyoil Kim is the corresponding author.}
%\thanks{This work was supported by Electronics and Telecommunications Research Institute(ETRI) grant funded by ICT R\&D program of MSIT/IITP[RS-2023-00217885, Development of integrating interference analysis technology for improving frequency utilization efficiency]}
\thanks{This work was supported by Institute of Information \& communications Technology Planning \& Evaluation (IITP) grant funded by the Korea government (MSIT) [RS-2023-00217885, Development of integrated interference analysis technology for improving frequency utilization efficiency].}
}

% The paper headers
%\markboth{submitted to IEEE Journal on Selected Areas in Communications}
%,~Vol.~14, No.~8, August~2021}%
%{}%Shell \MakeLowercase{\textit{et al.}}: A Sample Article Using IEEEtran.cls for IEEE Journals}

\maketitle

\begin{abstract}
\input{abstract}
\end{abstract}


\section{Introduction}
\label{sec:intro}
\input{introduction}

\section{System Model}
\label{sec:model}
\input{model}
\section{Proposed Markov Chain for AP MLD}
\label{sec:ap_mc}
\input{ap_mc}

\section{Proposed Markov Chain for Non-AP MLD}
\label{sec:nonap_mc}
\input{nonap_mc}

\section{Derivation of Key Parameters in the Proposed Markov Chains} %Performance Metrics %Analytical Model for Proposed Coexistence Scenario
\label{sec:analysis}
\input{analysis}

\section{Performance Evaluation}
\label{sec:evaluation}
\input{evaluation}

\section{Conclusion}
\label{sec:conclusion}
\input{conclusion}

\bibliographystyle{IEEEtran}
\bibliography{references}

\input{bios.tex}

\balance

\end{document}

%% file: abstract.tex
Wi-Fi 7 introduces Multi-link operation (MLO) to enhance throughput and latency performance compared to legacy Wi-Fi standards.
MLO enables simultaneous transmission and reception through multiple links, departing from conventional single-link operations (SLO).
To fully exploit MLO's potential, it is essential to %\textcolor{red}{develop effective MLO-based channel access mechanisms and} 
investigate Wi-Fi 7's coexistence performance with legacy Wi-Fi devices.
Existing approaches, however, have overlooked some crucial aspects of MLO, necessitating the development of a standards-compliant analytical framework to model the actual channel access mechanism of MLO.
Therefore, this paper tries to fill the gap by proposing a set of novel Markov chains (MC) to accurately model the MLO operation aligned with multi-link backoff behaviors specified by the standard.
Specifically, we design two separate MCs for AP and non-AP multi-link devices (MLD) respectively, based on which transmit and collision probabilities are derived under the saturated traffic condition.
Then, we also derive closed-form expressions for the throughput of various device types in the coexistence scenario between Wi-Fi 7 and legacy Wi-Fi, including AP MLD, non-AP MLD, and legacy devices.
To validate the accuracy of our proposed models, we developed an ns-3 based simulator by implementing both STR(simultaneous transmission and reception) and NSTR(non-STR) based MLO operations.
Our ns-3 based extensive simulations have demonstrated that the proposed analytic model provides accurate estimates on the per-device throughput performance, while also revealing the dynamics of inter-WLAN coexistence scenarios.
%\textcolor{blue}{Both analysis and simulation results demonstrate that, in coexistence under saturated conditions, legacy devices more frequently occupy the channel due to the complexity of MLO.}
%\textcolor{blue}{The analysis }
%Therefore, this research aims to addresses this gap by constructing an accurate Markov chain model for MLO aligned with the backoff behaviors specified by the standard, and by deriving closed-form expressions for the throughput of each device type in the Wi-Fi 7 and legacy Wi-Fi coexistence scenario.
%This research aims to deepen understanding regarding MLO's impact on Wi-Fi coexistence, providing a sturdy foundation for the advancement of wireless communication systems.

%% file: introduction.tex
%\textcolor{red}{\bf Note for SuHwan:}
%{\bf NOMS Doctoral Symposium requires us to:}
%\begin{itemize}
%\item clearly motivate, discuss, and summarize the proposed Ph.D. research,
%\item describe how the adopted approach fits into and advances research in related fields, and
%\item report on your progress. 
%\item In the concluding section, identify 2 questions/areas for improvement that you would like to discuss during the session.
%\end{itemize}

%\begin{itemize}	
%\item Wi-Fi 7, operating under the 802.11be standard, proposed Multi-Link Operation(MLO). 
%\\
%\item Research is needed to analyze the coexistence performance between MLO-based devices, denoted by multi-link devices (MLD), and legacy Wi-Fi devices.
%\\
%\item MLO necessitates consideration of start-time alignment and end-time alignment.
%\\
%\item Appl.Sci.'20 research conducted a mathematical study on MLD's coexistence performance, considering essential factors.
%\\
%\item However, Appl.Sci.'20 has overlooked certain aspects in representing the actual operation of MLD.
%\\
%\item In this research, we aim to construct Markov chains reflecting the actual operation of MLDs in coexistence to derive per-device throughput for each device type in a closed form.
%\\
%\item In \ref{sec:model}, we will address the concepts of start-time and end-time alignment and discuss the aspects overlooked by Appl.Sci.'20.
%\\
%\item In \ref{sec:progress}, we will propose solutions to address them.
%\end{itemize}

Wi-Fi 7, equivalently IEEE 802.11be, %EHT(Extremely High Throughput), 
introduces new features to achieve 
better performance than legacy Wi-Fi standards \cite{deng2020ieee}
and flexibility in utilizing unlicensed bands as agile as its competitors like NR-U \cite{Ssimbwa2024}.
Among them, one notable feature is Multi-Link Operation (MLO) which aims to achieve higher data rates and lower latency by introducing a new type of devices called Multi-Link Device (MLD) that can transmit and receive data simultaneously through multiple links \cite{lopez2022multi,khorov2020current}, unlike legacy Single-Link Devices (SLD) performing Single-Link Operation (SLO).
An MLD consists of multiple stations (STAs), each with a separate lower MAC but managed by the common upper MAC.
These affiliated STAs individually exchange data through their respective channels, allowing simultaneous transmission and reception (STR) \cite{murti2021multi, lopez2022dynamic}. 

STR based MLO, however, may not be possible for every MLD. 
While AP MLDs are considered capable of STR, non-AP MLDs may face challenges to perform STR due to IDC (In-Device Coexistence) interference caused by the proximity of transceivers on the same platform, i.e., transmission on one link self-interferes with reception on another link \cite{chen2022overview}.   %, leading to situations where STR cannot be performed.
Hence, non-AP MLDs adopt non-STR (NSTR) instead, which only allows simultaneous transmission or simultaneous reception on multiple links to resolve IDC interference \cite{adhikari2022analysis}. %issues and enables MLO. 
To support NSTR, the standard mandates alignment of start-times and/or %alignment of 
end-times for the transmissions across multiple links \cite{huang2022mutli}.

With the advent of MLO, it is essential to accurately predict the performance of MLDs when coexisting with legacy Wi-Fi devices (i.e., SLDs).
Several studies have addressed this issue using simulation-based approaches \cite{korolev2021analyses,korolev2022study,naribole2020simultaneous, carrascosa2023performance}.
\cite{korolev2021analyses} utilized an ns-3 based simulator to evaluate the throughput performance of each device type in a coexistence scenario involving an AP MLD operating in STR, a non-AP MLD operating in NSTR, and legacy Wi-Fi devices.
\cite{korolev2022study} also employed ns-3 simulations to investigate the impact of various EDCA-based channel access mechanisms on the throughput and latency performance in a similar coexistence scenario.
\cite{naribole2020simultaneous} used ns-3 as well, to measure the performance degradation caused by IDC interference when non-AP MLDs operate in STR, comparing it with their performance operating in NSTR.
In addition, \cite{carrascosa2023performance} evaluated the coexistence performance via simulations by focusing on how latency varies with traffic load. 
%\textcolor{red}{[HK: Is this simulation-based or analytical work?]}.

Simulation-based approaches, however, cannot outweigh an analytical framework due to the lack of generality and intuitions they can provide.
\textcolor{black}{Each simulation result is only tied to its specific topology and parameter setup, and sweeping through a meaningful range of scenarios could take hours or even days.
By contrast, a model-based analytical framework yields closed-form relations that instantly reveal how design knobs affect key performance metrics and provide insights that black-box simulation alone cannot offer.}
%\textcolor{blue}{Because the equations can be evaluated far more quickly than simulations, the framework can scan vast parameter spaces or even support real-time optimisation.}
%\textcolor{blue}{Equally important, the transparency of the mathematics provides insights that block-box simulation alone cannot offer.}
Unfortunately, only few have attempted to capture the behavior of MLDs using analytical models \cite{song2021performance, korolev2022analytical}.
%In such a vein, 
\cite{song2021performance} proposed a channel access scheme for NSTR in both uplink (UL) and downlink (DL) scenarios and derived the coexistence throughput performance in a closed form, but \textcolor{black}{its model forcibly consolidated all affiliated STAs in an MLD into a single combined Markov chain and thus misrepresents the standard-compliant alignment and backoff dynamics.}
%missing to consider some critical aspects regarding the legitimate MLO operations.
In the meantime, \cite{korolev2022analytical} developed an analytical model for NSTR operation in the presence of legacy devices with saturated traffic to estimate the throughput of NSTR MLDs. 
However, it did not account for start-time alignment, a key operation in MLO.

\textcolor{black}{To fill the gaps, we propose the first standards-compliant analytical framework that conforms with the per-STA backoff procedure and mandatory alignments, jointly models AP MLD and non-AP MLDs along with legacy SLDs, and accommodates asymmetric link configurations.
Specifically, this paper aims at constructing a set of novel Markov chains (MC) to reflect the correct MLO NSTR operations and to accurately derive per-device throughput for three types of coexisting devices -- AP MLD, non-AP MLDs, and SLDs.}
%Therefore,this paper aims at constructing a set of novel Markov chains (MC) to reflect the correct MLO NSTR operations (which have never been accurately modeled so far) and to accurately derive per-device throughput for three types of coexisting devices -- AP MLD, non-AP MLDs, and SLDs.
Our proposed MCs should play as a key building block for future MLO mechanisms to develop more intelligent and efficient MLO-based dynamic channel access ensuring the throughput and latency fairness among various types of coexisting WLAN devices.
\textcolor{black}{Note that this paper's preliminary conference version appeared at \cite{NOMS24} which only presented AP MLD's MC without providing a complete analysis (e.g., no derivation of stationary probabilities) nor simulation-based evaluations.
On the contrary, this journal manuscript provides a full set of MCs for both AP MLD and non-AP MLD with a series of complete analysis and evaluations, thus significantly extending the scope of its preliminary version.}

The paper is organized as follows. 
Section \ref{sec:model} overviews %the concept of 
start-time and end-time alignments, and discusses what aspects of them have been overlooked in the literature. %by \cite{song2021performance}.
Section \ref{sec:ap_mc} introduces our proposed MC for an AP MLD, and derives its stationary and transmit probabilities.
%Then, Section \ref{sec:ap_mc} introduces our proposed MC to address the issues for AP MLDs, and derives three types of per-device throughput.
Similarly, Section \ref{sec:nonap_mc} proposes our novel MC for Non-AP MLDs along with the same set of derivations. 
Next, Section \ref{sec:analysis} introduces the derivation of other key parameters used in the proposed analytical framework.
%Afterwards, Section \ref{sec:evaluation} presents extensive simulation results to show the accuracy of the proposed MCs in analyzing the coexistence performance. 
Then, Section \ref{sec:evaluation} presents ns-3 based simulations compared to the proposed analysis, showing the efficacy of our proposal along with some key lessons on the coexistence dynamics.
Finally, the paper concludes with Section~\ref{sec:conclusion}.

%% file: model.tex
\begin{figure}[!t]
%	\centerline{\includegraphics[width=0.65\linewidth]{figures/Fig1_system_model.eps}}
	\centerline{\includegraphics[width=0.65\linewidth]{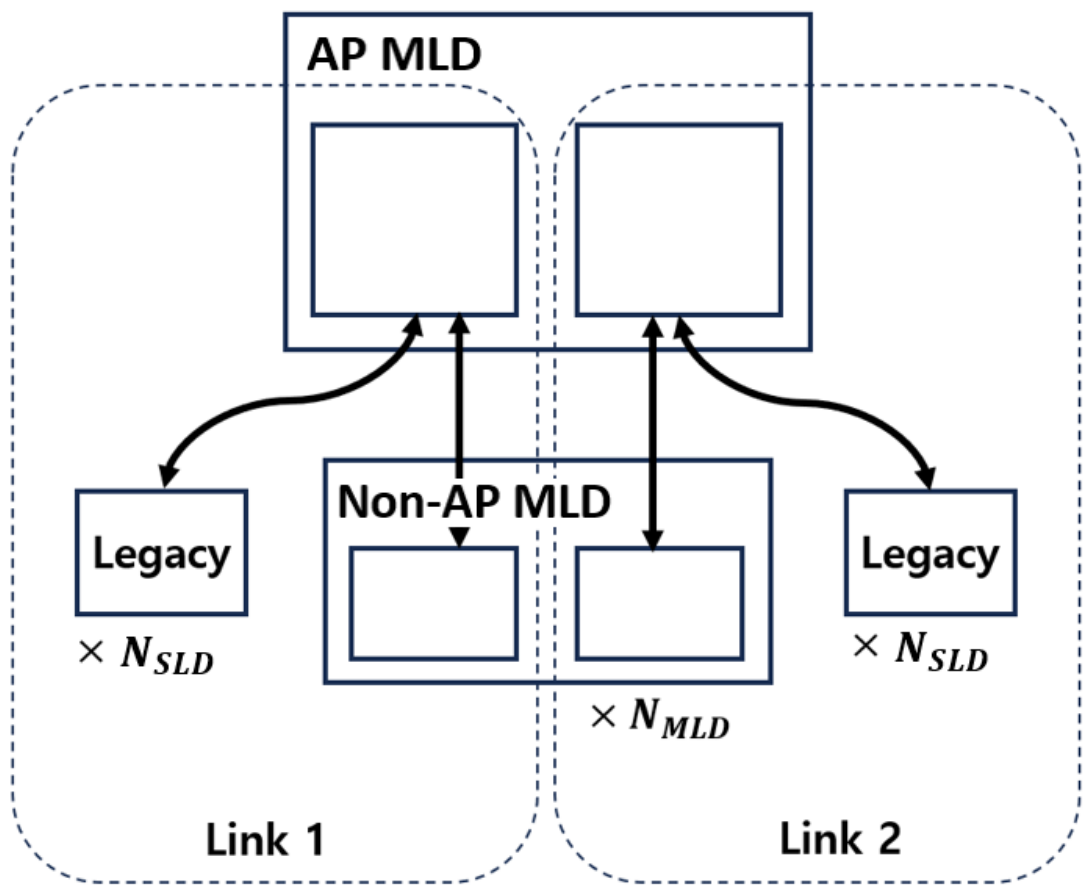}}
%	\centerline{\includegraphics[width=0.65\linewidth]{Fig.1_system_model.eps}}
	\vspace{-0.05in}
	\caption{Coexistence scenario between MLDs and legacy devices}
	\label{fig:architecture}
	\vspace{-0.05in}
\end{figure}

\begin{figure}[!t]
%	\centerline{\includegraphics[width=0.8\linewidth]{figures/Fig2_end_time_alignment.eps}}
	\centerline{\includegraphics[width=0.8\linewidth]{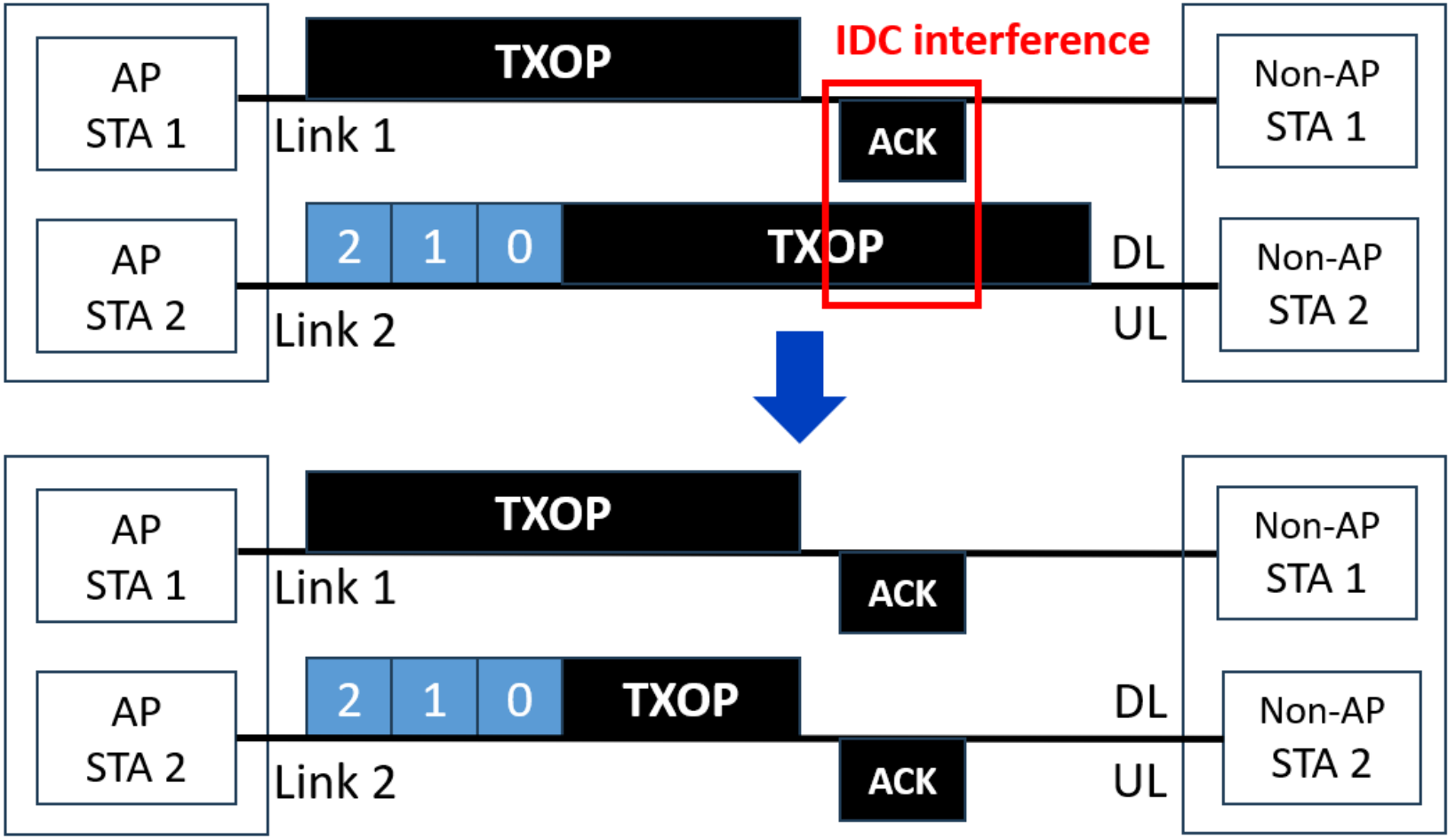}}
	\vspace{-0.05in}
	\caption{Resolving IDC interference through end-time alignment}
	\label{fig:endtime}
	\vspace{-0.15in}
\end{figure}

\begin{figure}[!t]
%	\centerline{\includegraphics[width=0.8\linewidth]{figures/Fig3_start_time_alignment.eps}}
	\centerline{\includegraphics[width=0.8\linewidth]{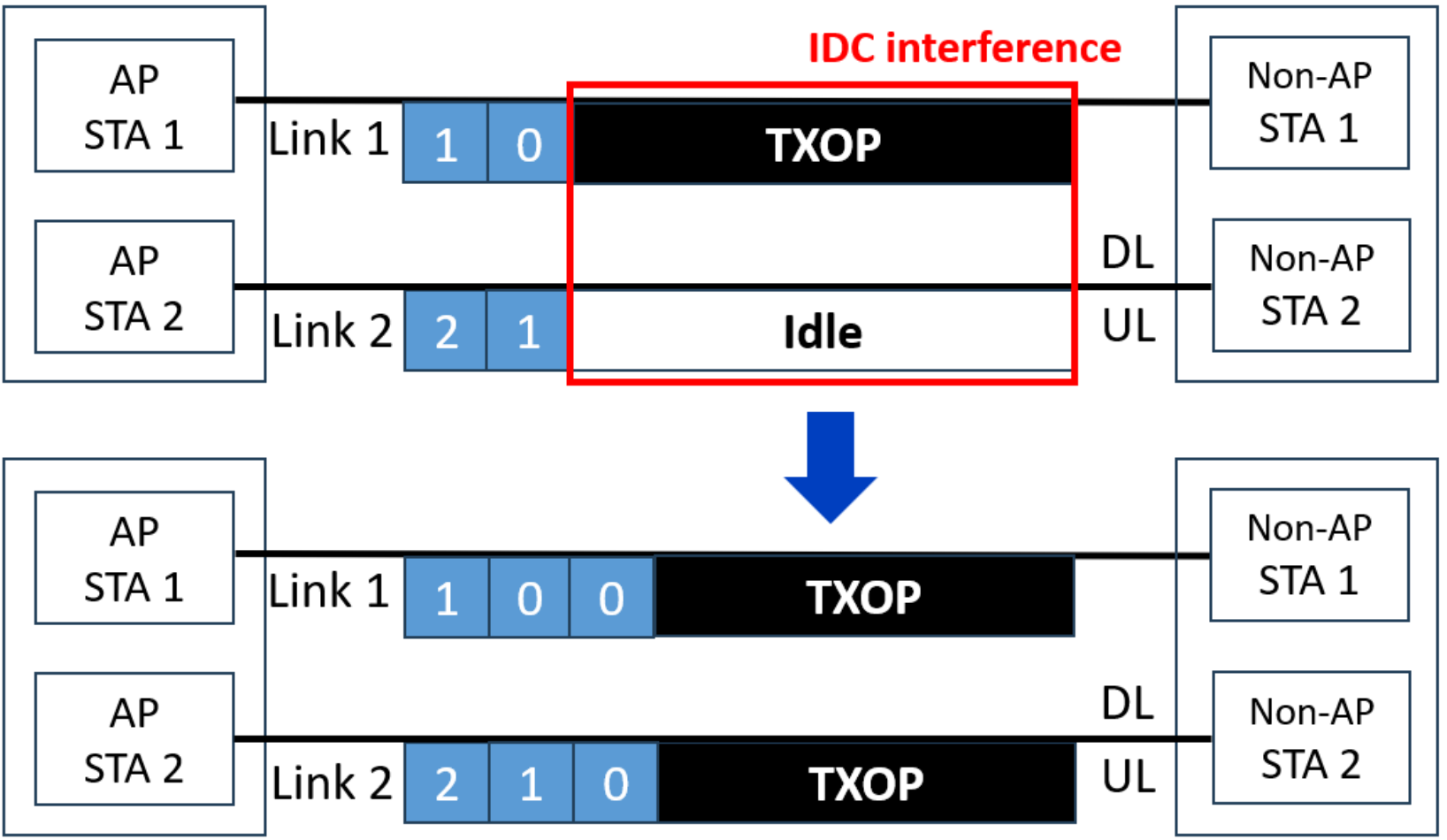}}
	\vspace{-0.05in}
	\caption{Resolving IDC interference through start-time alignment}
	\label{fig:starttime}
	\vspace{-0.15in}
\end{figure}

\begin{comment}
\begin{figure}[!t]
%	\centerline{\includegraphics[width=0.8\linewidth]{figures/Fig4_correct_backoff_operation.eps}}
	\centerline{\includegraphics[width=0.8\linewidth]{figures/Fig4_correct_backoff_operation-eps-converted-to.pdf}}
	\caption{Standard-compliant backoff operation during DL transmission}
	\label{fig:correct}
	\vspace{-0.1in}
\end{figure}
\end{comment}

We consider a WLAN coexistence scenario shown in Fig.~\ref{fig:architecture}, which includes an AP MLD, $N_{MLD}$ non-AP MLDs, and $N_{SLD}$ 802.11ax devices, where each MLD consists of two STAs to access two links.
%The AP and non-AP devices communicate through two links, assuming the presence of $N_{legacy}$ devices and $N_{MLD}$ STAs for each link.
We assume a saturation traffic condition as in \cite{bianchi2000performance}.
%The system is slotted under saturation condition \cite{bianchi2000performance}.
%As mentioned in Sec \ref{sec:intro}, 
The AP MLD can perform STR, whereas non-AP MLDs need to operate based on NSTR. %to avoid IDC interference.
%IDC interference occurring on one link within the MLD can disrupt reception or channel sensing on other links. 

%\textcolor{red}{According to \cite{magrin2023performance}, legacy devices are assumed to access the medium following a Bianchi model-based MC.}

%The impact of how each type of MLD accesses the medium for DL and UL communication is as follows. 
\textcolor{black}{In accordance with the standard, DL and UL transmissions should selectively apply end-time alignment and start-time alignment} as follows.
First, when the AP MLD performs DL transmission, it must conduct {\it end-time alignment} as shown in Fig.~\ref{fig:endtime}.
This is because DL transmission on one link incurs an ACK frame transmitted by the receiving non-AP MLD, thus inducing IDC interference on another link unless the end-times of the two transmissions are aligned.
%End-time alignment can resolve this issue by synchronizing the ending time of DL transmissions on multiple links.
%
Next, when a non-AP MLD performs UL transmission, both end-time alignment and {\it start-time alignment} must be conducted where the latter is shown in Fig.~\ref{fig:starttime}, because UL transmission on one link interferes with the clear channel assessment (CCA) on another link due to IDC interference. 
That is, even though the other link is actually idle, it is incorrectly perceived as busy. %, causing them not to initiate transmission.
Start-time alignment resolves this problem by synchronizing the initiation of UL transmissions across multiple links, while end-time alignment should be adopted together to prevent ACKs from the AP MLD from causing IDC interference.

\textcolor{black}{Building on such standard-compliant behaviors, we consider following AP-MLD channel access rules to avoid IDC interference.
When the backoff counter of one link (say Link 1, without loss of generality) in the AP MLD reaches 0, its action depends on the state of the other link (say Link 2, without loss of generality) belonging to the same MLD, as follows:
\begin{itemize}
%\begin{enumerate}[leftmargin=*]
\small 
\item If Link 2 senses its channel idle, Link 1 transmits immediately. %the zero-counter link
\item If Link 2 senses its channel busy without transmitting itself, Link 1 restarts its backoff procedure.
\item If Link 2 is already transmitting a DL frame to a destination different from Link 1's, Link 1 transmits immediately.
\item If Link 2 is already transmitting a DL frame to the same destination as Link 1's, Link 1 performs end-time alignment.
%\end{enumerate}
\end{itemize}
}

\textcolor{black}{Non-AP MLDs follow a procedure different from that of the AP MLD.
When Link 1's backoff counter reaches zero, it freezes until Link 2's backoff counter in the same non-AP MLD also reaches zero.
If either channel becomes busy during the wait, Link 1 immediately restarts its backoff process.
Once both counters reach zero, %with the channels idle, 
the non-AP MLD triggers a simultaneous uplink transmission on both links, applying start-time alignment as well as end-time alignment.
Note that the aforementioned channel access rules 
reflect a most conservative interpretation 
%\textcolor{red}{embody a deliberately conservative reading} 
of the standard, to ensure that 
they are fully standard-compliant.}
%all start-time and end-time alignment actions remain fully standard-compliant.}

\cite{song2021performance} proposed backoff schemes and the corresponding MCs for AP and non-AP MLDs, taking into account the aforementioned factors. 
However, its MCs have the following limitations. 
First, they failed to accurately represent the backoff behavior of the standard-compliant MLDs.
Second, each STA within an MLD independently accesses its own channel, and thus each STA's behavior should be individually modeled by a separate MC.
However, \cite{song2021performance} over-simplified it into a single `combined' MC.

In Sections \ref{sec:ap_mc} and \ref{sec:nonap_mc}, we will first elaborate legitimate backoff behaviors of MLDs that have been overlooked by \cite{song2021performance}, %properly represented 
%and then propose our novel MCs that can accurately incorporate them. 
\textcolor{black}{and then propose our novel MCs that not only capture these behaviors but also model each link's channel access independently.
To this end, we also present a detailed analysis on our MCs, including the new states and transitions required to accurately model standard-compliant MLO behaviors.}
Based on the proposed architectures, we will also formulate per-device per-link throughput in a closed form, inspired by the methods proposed in \cite{bianchi2000performance} and \cite{magrin2023performance}.
In particular, Section \ref{sec:ap_mc} will propose our MC for AP MLDs, while Section \ref{sec:nonap_mc} will focus on our MC for non-AP MLDs.

%% file: ap_mc.tex
%\begin{figure}
%	\centerline{\includegraphics[width=0.9\linewidth]{figures/Fig.4_end_time_alignment_Appl.eps}}
%	\caption{\textcolor{blue}{Multi-link access with end-time alignment proposed in \cite{song2021performance}}}
%	\label{fig:temp}
%	\vspace{-0.1in}
%\end{figure}

%\begin{comment}
\begin{figure}[!t]
%	\centerline{\includegraphics[width=0.8\linewidth]{figures/Fig4_correct_backoff_operation.eps}}
	\centerline{\includegraphics[width=0.8\linewidth]{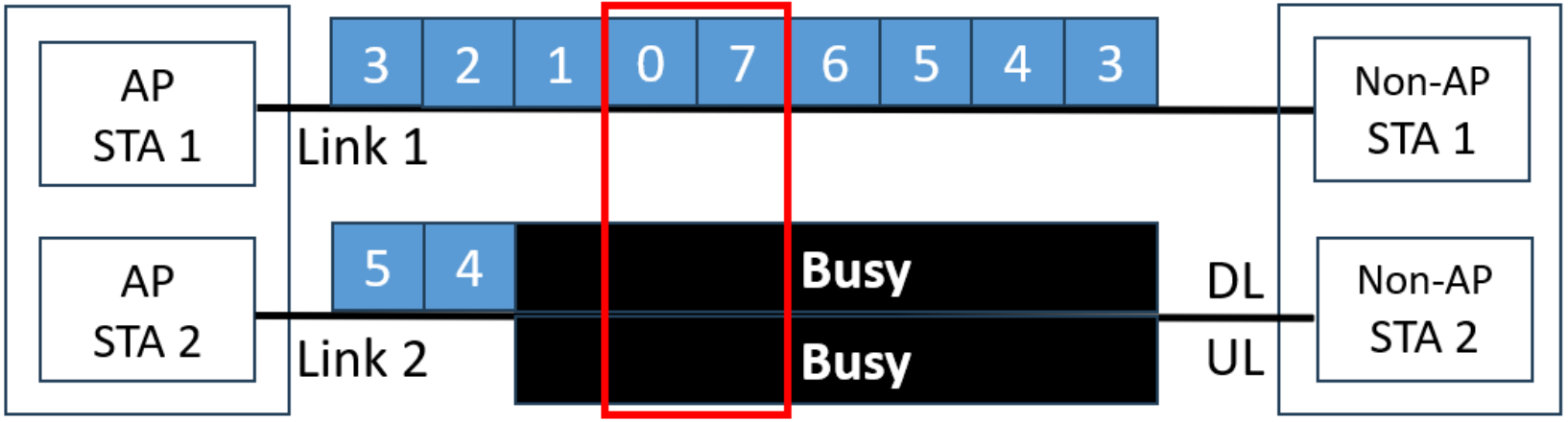}}
	\caption{Standard-compliant backoff operation during DL transmission}
	\label{fig:correct}
	\vspace{-0.1in}
\end{figure}
%\end{comment}

\begin{figure}[!ht]
%	\centerline{\includegraphics[width=0.99\linewidth]{figures/Fig5_distinction_state_0.eps}}
	\centerline{\includegraphics[width=0.99\linewidth]{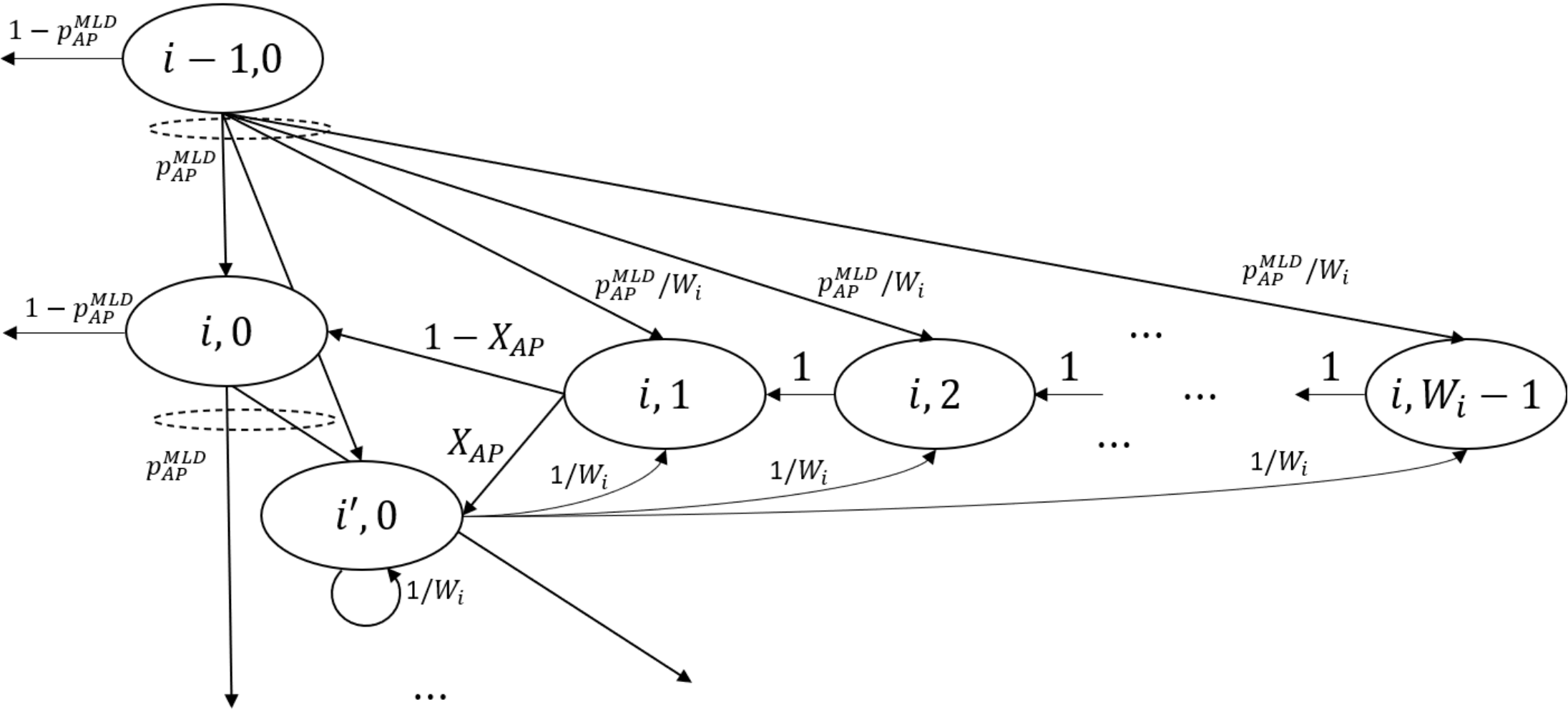}}
	\caption{Separation of state $(i,0)$ into $(i,0)$ and $(i',0)$}
	\label{fig:distinction}
	\vspace{-0.1in}
\end{figure}

\begin{figure*}[!t]
%	\centerline{\includegraphics[width=0.85\linewidth]{figures/Fig6_combined_MC_framework_for_AP_MLD.eps}}
	\centerline{\includegraphics[width=0.85\linewidth]{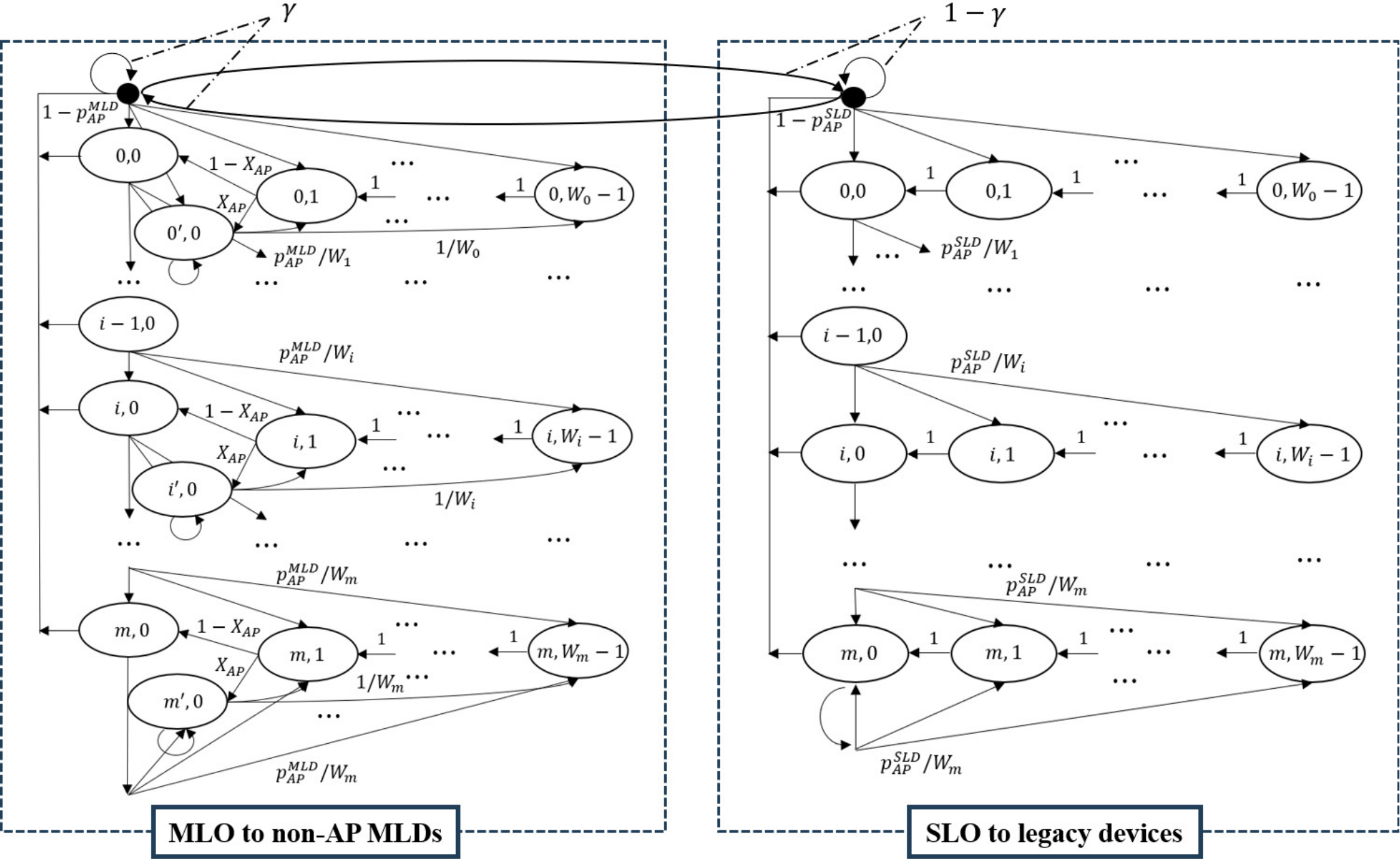}}
	\caption{The proposed MC for the AP MLD (left-hand side: DL to non-AP MLDs, right-hand side: DL to legacy devices)} %\textcolor{red}{[HK: stage $m$'s states have typos: $i \rightarrow m$][HK: is it ok to abuse $(0,0)$, $(0,1)$, $(i,0)$, etc., on both sides?]}}
	\label{fig:combined}
	\vspace{-0.1in}
\end{figure*}

We first explain the discrepancy between the backoff scheme in the Wi-Fi 7 standard and the one modeled in \cite{song2021performance}.
According to the standard, in case the AP MLD's backoff counter reaches 0 in Link 1 but Link 2 is busy at the moment not because of the AP's transmission, the AP MLD should restart the backoff in Link 1 while retaining the current backoff stage, as depicted in Fig. \ref{fig:correct}.
The model in \cite{song2021performance}, however, failed to capture this behavior, which would incur significant analytical errors especially at high traffic load where channels become busy quite frequently.
%\cite{song2021performance} assumed that once an idle channel is discovered it remains idle for an extended period of time, which is certainly not true since a channel can become busy anytime by the transmission from other devices. 
%In saturated conditions where most channels are busy, this would happen frequently, limiting MLD's chances to commence transmission.

To address this issue in our MC, we introduce a state transition that restarts the backoff when the backoff counter reaches 0 in the aforementioned scenario.
Unlike in the original Bianchi MC where transmission is mandatory upon reaching a backoff counter of 0, with our newly added transition, there is now an option to restart the backoff instead of forcing transmission if the other channel is busy.
Fig. \ref{fig:distinction} illustrates this clearly, where the MC's state $(i,k)$ represents the backoff stage $i$ and the backoff counter $k$, with $0 \le i \le m$ and $0 \le k \le W_i-1$. 
Specifically, there exist two transitions from state $(i,1)$ into either state $(i,0)$ or state $(i',0)$, each corresponding to actual transmission and restarted backoff, respectively.
In the figure, $X$ represents the probability that the other link is busy not due to the AP's transmission. 

Next, we build MCs for individual STAs belonging to the AP MLD. 
\cite{song2021performance} assumed that the backoff of STAs within an MLD occurs simultaneously, based on which it oversimplified each STA's individual MC into a single combined MC.
We rectify this by introducing a separate MC per STA, ensuring that each STA within the AP MLD independently performs  backoff based on its own link state, while incorporating our new state transition proposed.

Finally, we propose a novel framework to describe the correct behavior of the AP MLD. 
The channel access of the AP MLD is categorized into two cases based on the destination of the packet that the AP MLD tries to transmit, as follows. 
\begin{itemize}
	\item Case 1: If the packet's destination is a legacy device (with probability $1-\gamma$), it is transmitted via a single link following the traditional single link operation (SLO). %following the channel access method of IEEE 802.11ax.
	\item Case 2: If the packet's destination is a non-AP MLD (with probability $\gamma$), MLO is performed by following the AP MLD's MC we propose.
\end{itemize}
Building upon the aforementioned enhancements, our novel MC for the AP MLD is constructed as depicted in Fig. \ref{fig:combined}.

Utilizing our proposed MC and the methodologies introduced in \cite{bianchi2000performance} and \cite{magrin2023performance}, we formulate per-device per-link throughput in a closed form. 
Initially, we determine the {\it transmit} probability $\tau$ that a STA transmits in a randomly chosen slot, according to: the proposed MC for the AP MLD, the MC in \cite{song2021performance} for non-AP MLDs,\footnote{Note that we plan to apply the proposed ideas in this paper to the MC of non-AP MLDs, as our future work.} and the MC in \cite{bianchi2000performance} for legacy STAs.
We denote $\tau$ of each device type by $\tau_{\text{AP}}^{\text{SLD}}$, $\tau_{\text{MLD}}^I$, $\tau_{\text{SLD}}^I$ for Case 1, and by $\tau_{\text{AP}}^{\text{MLD}}$, $\tau_{\text{MLD}}^{II}$, $\tau_{\text{SLD}}^{II}$ for Case 2. 
From $\tau$, we can derive the {\it collision} probability $p$ that a device would experience a collision in a randomly chosen slot, where $p$ for each device type is denoted by $p_{\text{AP}}^{\text{SLD}}$, $p_{\text{MLD}}^I$, $p_{\text{SLD}}^I$ for Case 1, and by $p_{\text{AP}}^{\text{MLD}}$, $p_{\text{MLD}}^{II}$, $p_{\text{SLD}}^{II}$ for Case 2.

In the sequel, we present the derivation of $\tau_{\text{AP}}^{\text{SLD}}$ and $\tau_{\text{AP}}^{\text{MLD}}$.
In the meantime, the derivation of $\tau_{\text{MLD}}^I$ and $\tau_{\text{MLD}}^{II}$ will be presented in Section~\ref{sec:nonap_mc}.
On the other hand, the derivation of $\tau_{\text{SLD}}^I$ and $\tau_{\text{SLD}}^{II}$ closely follows the one in \cite{bianchi2000performance}, which will be presented in Section \ref{subsec:tau_leagcy}.
After deriving all sorts of $\tau$'s, various types of $p$'s will be derived in Section \ref{subsec:collision_prob}.

%--------------------------------------------------------------
\subsection{Derivation of stationary probabilities} %transmit probabilities} %$\tau_{AP}$}}
We first derive the stationary probabilities of the AP MLD's MC in Fig. \ref{fig:combined}, which consists of the MLO part (on the left) and the SLO part (on the right). 
The stationary probabilities are denoted by $b_{i,k}^{\text{MLD}}$ if corresponding to the MLO part, or by $b_{i,k}^{\text{SLD}}$ if corresponding to the SLO part.
After deriving $b_{i,k}^{\text{MLD}}$ and $b_{i,k}^{\text{SLD}}$, we will present how to derive the two transmit probabilities $\tau_{AP}^{SLD}$ and $\tau_{AP}^{MLD}$.

First, the vertical state transitions in the MLO part among its leftmost states $(i,0)$, $0 \le i \le m$ provide
\begin{align}
b_{i-1,0}^{\text{MLD}} \cdot p_{\text{AP}}^{\text{MLD}} &= b_{i,0}^{\text{MLD}}, \quad 0 < i < m  \notag \\
	\Rightarrow b_{i,0}^{\text{MLD}} &= \left( p_{\text{AP}}^{\text{MLD}} \right)^i \cdot b_{0,0}^{\text{MLD}}, \quad 0 \leq i < m , \label{eq:b_MLD_1} \\
b_{m-1,0}^{\text{MLD}} \cdot p_{\text{AP}}^{\text{MLD}} &= (1 - p_{\text{AP}}^{\text{MLD}}) \cdot b_{m,0}^{\text{MLD}}  \notag \\
	\Rightarrow b_{m,0}^{\text{MLD}} &= \frac{p_{\text{AP}}^{\text{MLD}}}{1 - p_{\text{AP}}^{\text{MLD}}} \cdot b_{m-1,0}^{\text{MLD}} 
    	= \frac{(p_{\text{AP}}^{\text{MLD}})^m}{1 - p_{\text{AP}}^{\text{MLD}}} \cdot b_{0,0}^{\text{MLD}} . \label{eq:b_MLD_2} 
\end{align}
By Eqs. \eqref{eq:b_MLD_1} and \eqref{eq:b_MLD_2}, we obtain
\begin{align}
\sum_{i=0}^{m} b_{i,0}^{\text{MLD}} 
%&= \sum_{i=0}^{m-1} b_{i,0}^{\text{MLD}} 
%	+ \frac{(p_{\text{AP}}^{\text{MLD}})^m}{1 - p_{\text{AP}}^{\text{MLD}}} \cdot b_{0,0}^{\text{MLD}} \notag \\
%&= 
 = \left[ 
 	\frac{1 - (p_{\text{AP}}^{\text{MLD}})^m}{1 - p_{\text{AP}}^{\text{MLD}}} 
	+ \frac{(p_{\text{AP}}^{\text{MLD}})^m}{1 - p_{\text{AP}}^{\text{MLD}}} 
   \right] b_{0,0}^{\text{MLD}} 
   = \frac{b_{0,0}^{\text{MLD}}}{1 - p_{\text{AP}}^{\text{MLD}}} . \label{eq:(3)} 
	%\quad \textcolor{red}{\text{... (3)}} 
\end{align}
Similarly, the vertical state transitions in the SLO part among its leftmost states $(i,0)$, $0 \le i \le m$ lead to
\begin{align}
\sum_{i=0}^{m} b_{i,0}^{\text{SLD}} &= \frac{b_{0,0}^{\text{SLD}}}{1 - p_{\text{AP}}^{\text{SLD}}} . \label{eq:(4)} 
	%\quad \textcolor{red}{\text{... (4)}}
\end{align}

\begin{comment}
At stage $1, 2, \ldots, m$, the proposed AP-MLD MC is consistent with the original Bianchi MC \cite{?}, while the difference lies at stage 0.
\textcolor{red}{[it remains consistent with the existing modified AP-MLD M.C. and the original Bianchi M.C., with the difference being at stage 0.]}
\end{comment}

Next, let us consider the states with $1 \leq k \leq W_0 - 1$. 
For the MLO part, when $i = 0$ and $1 \leq k \leq W_0 - 1$, we obtain
\begin{align}
b_{0,k}^{\text{MLD}} &= \frac{W_0 - k}{W_0} \left[ \gamma ( b_{0,0}^{\text{MLD}} + b_{0,0}^{\text{SLD}} ) + b_{0',0}^{\text{MLD}} \right] , 
	1 \leq k \leq W_0 - 1 , \label{eq:(5)} %\\
%&\quad \textcolor{red}{... (5)} \notag
\end{align}
\textcolor{black}{by applying generalized global balance equations (g.g.b.e.) to each $b_{0,k}^{\text{MLD}}$.}
\begin{comment}
\begin{align}
1 \cdot b_{0,W_0 - 1}^{\text{MLD}} =& \phantom{0} \gamma \cdot \left\{ (1 - p_{\text{AP}}^{\text{MLD}}) \sum_{i=0}^{m} b_{i,0}^{\text{MLD}} \right. \notag \\
	&+ \left. (1 - p_{\text{AP}}^{\text{SLD}}) \sum_{i=0}^{m} b_{i,0}^{\text{SLD}} \right\} \cdot \frac{1}{W_0} 
		+ \frac{1}{W_0} \cdot b_{0',0}^{\text{MLD}} \notag \\
=& \phantom{0} \frac{1}{W_0} \left[ \gamma \left( b_{0,0}^{\text{MLD}} + b_{0,0}^{\text{SLD}} \right) + b_{0',0}^{\text{MLD}} \right] , \notag%\\
\end{align}
\begin{align}
1 \cdot b_{0,W_0 - 2}^{\text{MLD}} =& \phantom{0} \frac{1}{W_0} \left[ \gamma \left( b_{0,0}^{\text{MLD}} + b_{0,0}^{\text{SLD}} \right) + b_{0',0}^{\text{MLD}} \right] + 1 \cdot b_{0,W_0 - 1}^{\text{MLD}} \notag \\
	=& \phantom{0} \frac{2}{W_0} \left[ \gamma \left( b_{0,0}^{\text{MLD}} + b_{0,0}^{\text{SLD}} \right) + b_{0',0}^{\text{MLD}} \right] , \notag\\
&\qquad \qquad \vdots \notag\\
1 \cdot b_{0,1}^{\text{MLD}} =& \phantom{0} \frac{W_0 - 1}{W_0} \left[ \gamma \left( b_{0,0}^{\text{MLD}} + b_{0,0}^{\text{SLD}} \right) + b_{0',0}^{\text{MLD}} \right] , \notag
\end{align}
which gives
\end{comment} 

\textcolor{black}{Building on the standard-compliant behavior described in Section \ref{sec:model}, once one link's backoff counter in an AP MLD reaches zero, its next action depends on the ongoing transmission on or the channel state of the other link.
To quantify such influence between links as part of the MC, let us define $X_{\text{AP}}$ as `the probability that from the viewpoint of a certain link belonging to the AP MLD, the other link is busy due to the transmission either from non-AP MLD or legacy STA'.
Note that $X_{\text{AP}}$ will be derived in Section \ref{subsec:derivation_x_y} since it requires derivation of $\tau$'s and $p$'s.} %\footnote{Deriving $X_{AP}$ requires $\tau$'s and $p$'s, and thus it will be presented in Section \ref{subsec:derivation_x_y} following the derivation of $\tau$'s and $p$'s.}
Then, by considering the state transitions from/to state $(0',0)$ of the MLO part, we have
\begin{align}
b_{0',0}^{\text{MLD}} &= X_{\text{AP}} \cdot \frac{1}{W_0} \cdot \gamma \cdot (b_{0,0}^{\text{MLD}} + b_{0,0}^{\text{SLD}}) + X_{\text{AP}} \cdot b_{0,1}^{\text{MLD}} \notag \\
&\quad + \frac{1}{W_0} \cdot b_{0',0}^{\text{MLD}} ,
%b_{0',0}^{\text{MLD}} = X_{\text{AP}} \cdot \frac{1}{W_0} \cdot \gamma \cdot %(b_{0,0}^{\text{MLD}} + b_{0,0}^{\text{SLD}}) 
%+ X_{\text{AP}} \cdot b_{0,1}^{\text{MLD}} + \frac{1}{W_0} \cdot b_{0',0}^{\text{MLD}} , %\notag %\\	
\end{align}
Applying Eq. \eqref{eq:(5)} to the above gives
\begin{align}
%\frac{W_0 - 1}{W_0} \cdot b_{0',0}^{\text{MLD}} &= X \cdot \frac{1}{W_0} \cdot \gamma \cdot (b_{0,0}^{\text{MLD}} + b_{0,0}^{\text{SLD}}) \notag \\
%	&\quad+ X \cdot \frac{W_0 - 1}{W_0} \left[ \gamma \cdot (b_{0,0}^{\text{MLD}} + b_{0,0}^{\text{SLD}}) + b_{0',0}^{\text{MLD}} \right]  \quad \notag \\ %\textcolor{red}{\text{by (5)}} \notag \\
%	&= X \cdot \gamma \cdot (b_{0,0}^{\text{MLD}} + b_{0,0}^{\text{SLD}}) %\notag \\
	%&\quad
%	+ X \cdot \frac{W_0 - 1}{W_0} \cdot b_{0',0}^{\text{MLD}} , \notag \\
%--------------
%\end{align}
%which instantly gives
%\begin{align}
%(1 - X) \cdot \frac{W_0 - 1}{W_0} \cdot b_{0',0}^{\text{MLD}} = X \cdot \gamma \cdot (b_{0,0}^{\text{MLD}} + b_{0,0}^{\text{SLD}}) . \notag %\\
%\end{align}
%Therefore, we obtain 
%\begin{align}
%--------------
%\therefore 
b_{0',0}^{\text{MLD}} &= \frac{X_{\text{AP}}}{1 - X_{\text{AP}}} \cdot \frac{W_0}{W_0 - 1} \cdot \gamma \cdot (b_{0,0}^{\text{MLD}} + b_{0,0}^{\text{SLD}}) . 
	%\quad \textcolor{red}{\text{... (6)}} 
	\label{eq:(6)}
\end{align}
%Then, by considering 
By the state transitions from/to state $(0,0)$ of the MLO part, 
\begin{align}
b_{0,0}^{\text{MLD}} 
	&= \frac{1 - X_{\text{AP}}}{W_0} \cdot \gamma \cdot (b_{0,0}^{\text{MLD}} + b_{0,0}^{\text{SLD}}) 
		+ (1 - X_{\text{AP}}) \cdot b_{0,1}^{\text{MLD}} \notag \\
%b_{0,0}^{\text{MLD}} 
%&= (1 - X_{\text{AP}}) \cdot \frac{1}{W_0} \cdot \gamma \cdot (b_{0,0}^{\text{MLD}} + b_{0,0}^{\text{SLD}}) 
%+ (1 - X_{\text{AP}}) \cdot b_{0,1}^{\text{MLD}} \notag \\		
%	&= (1 - X) \cdot \frac{1}{W_0} \cdot \gamma \cdot (b_{0,0}^{\text{MLD}} + b_{0,0}^{\text{SLD}})  \notag \\
%	&\quad+ (1 - X) \cdot \frac{W_0 - 1}{W_0} \left[ \gamma \cdot (b_{0,0}^{\text{MLD}} + b_{0,0}^{\text{SLD}}) + b_{0',0}^{\text{MLD}} \right] \quad  \notag \\
%	&= (1 - X) \gamma \cdot (b_{0,0}^{\text{MLD}} + b_{0,0}^{\text{SLD}}) 
%		+ (1 - X) \frac{W_0 - 1}{W_0} \cdot b_{0',0}^{\text{MLD}} \notag \\
%	&= (1 - X) \gamma \cdot (b_{0,0}^{\text{MLD}} + b_{0,0}^{\text{SLD}}) \notag \\
%	&\quad+ (1 - X) \frac{W_0 - 1}{W_0} \cdot \frac{X}{1 - X} \frac{W_0}{W_0 - 1} \gamma (b_{0,0}^{\text{MLD}} + b_{0,0}^{\text{SLD}}) \quad \notag \\
%	&= (1 - X) \gamma \cdot (b_{0,0}^{\text{MLD}} + b_{0,0}^{\text{SLD}}) 
%		+ X \gamma \cdot (b_{0,0}^{\text{MLD}} + b_{0,0}^{\text{SLD}}) \notag \\
	&= \gamma \cdot (b_{0,0}^{\text{MLD}} + b_{0,0}^{\text{SLD}}) , 
\end{align}
where we have applied Eqs. \eqref{eq:(5)} and \eqref{eq:(6)}. %to the second and fourth equalities, respectively.
As a result, we obtain 
\begin{align}
(1 - \gamma) \cdot b_{0,0}^{\text{MLD}} &= \gamma \cdot b_{0,0}^{\text{SLD}} %\quad \textcolor{red}{\text{... (7)}} 
	. \label{eq:(7)} %\\
%\Rightarrow \quad b_{0,0}^{\text{MLD}} &= \frac{\gamma}{1 - \gamma} \cdot b_{0,0}^{\text{SLD}} \notag \\
%b_{0,0}^{\text{SLD}} &= \frac{1 - \gamma}{\gamma} \cdot b_{0,0}^{\text{MLD}} \notag \\
%\Leftrightarrow \quad b_{0,0}^{\text{MLD}} + b_{0,0}^{\text{SLD}} &= \frac{b_{0,0}^{\text{MLD}}}{\gamma} \quad \textcolor{red}{\text{... (8)}} \label{eq:(8)}
\end{align}
By Eqs. \eqref{eq:(6)} and \eqref{eq:(7)},
\begin{align}
b_{0',0}^{\text{MLD}} &= \frac{X_{\text{AP}}}{1 - X_{\text{AP}}} \cdot \frac{W_0}{W_0 - 1} \cdot b_{0,0}^{\text{MLD}} .
	%\quad \textcolor{red}{\text{... (9)}} 
	\label{eq:(9)}
\end{align}
By combining Eqs. \eqref{eq:(5)}, \eqref{eq:(7)}, and \eqref{eq:(9)}, we obtain
\begin{align}
b_{0,k}^{\text{MLD}} 
%&= \frac{W_0 - k}{W_0} \cdot \left( b_{0,0}^{\text{MLD}} + b_{0',0}^{\text{MLD}} \right) , \quad 1 \le k \le W_0 - 1, \notag \\
&= \frac{W_0 - k}{W_0} \left[ 1 + \frac{X_{\text{AP}}}{1 - X_{\text{AP}}} \cdot \frac{W_0}{W_0 - 1} \right] b_{0,0}^{\text{MLD}} , \notag \\
&\qquad\qquad\qquad\qquad\qquad\qquad 1 \le k \le W_0 - 1,
	%\quad \textcolor{red}{\text{... (11)}} 
	\label{eq:(11)}
%b_{0,0}^{\text{MLD}} &= b_{0,0}^{\text{MLD}} \quad , k = 0 
\end{align}

For the SLO part, when $i = 0$ and $1 \leq k \leq W_0 - 1$, by Eqs. \eqref{eq:(3)} and \eqref{eq:(4)} and following the similar procedure shown earlier for the MLO part, we obtain
\begin{comment}
\textcolor{magenta}{	%%%%%%%%%%%%%%%%%%%%%%%%
\begin{align}
1 \cdot b_{0,W_0 - 1}^{\text{SLD}} &= \frac{1}{W_0} \cdot (1 - \gamma) \cdot \left[ (1 - p_{\text{AP}}^{\text{SLD}}) \sum_{i=0}^{m} b_{i,0}^{\text{SLD}} \right. \notag \\
	&\qquad \qquad \qquad \qquad \left. + (1 - p_{\text{AP}}^{\text{MLD}}) \sum_{i=0}^{m} b_{i,0}^{\text{MLD}} \right] \notag \\
	&= \frac{1}{W_0} \cdot (1 - \gamma) \cdot \left( b_{0,0}^{\text{SLD}} + b_{0,0}^{\text{MLD}} \right) , \notag \\
1 \cdot b_{0,W_0 - 2}^{\text{SLD}} &= \frac{1}{W_0} \cdot (1 - \gamma) \cdot \left( b_{0,0}^{\text{SLD}} + b_{0,0}^{\text{MLD}} \right) + 1 \cdot b_{0,W_0 - 1}^{\text{SLD}} , \notag \\
	&= \frac{2}{W_0} \cdot (1 - \gamma) \cdot \left( b_{0,0}^{\text{SLD}} + b_{0,0}^{\text{MLD}} \right) \notag \\
	&\quad \vdots \notag \\
1 \cdot b_{0,1}^{\text{SLD}} &= \frac{W_0 - 1}{W_0} \cdot (1 - \gamma) \cdot \left( b_{0,0}^{\text{SLD}} + b_{0,0}^{\text{MLD}} \right) , \notag 
\end{align}
from which we have
}	%%%%%%%%%%%%%%%%%%%%%%%%%%%%%%%%%%%%%%%%%%%%
\end{comment}
\begin{align}
b_{0,k}^{\text{SLD}} &= \frac{W_0 - k}{W_0} 
	(1 - \gamma) 
	\left( b_{0,0}^{\text{SLD}} + b_{0,0}^{\text{MLD}} \right) , 
	\; 1 \leq k \leq W_0 - 1 . \label{eq:b_0k_legacy}
\end{align}
Then, applying Eq. \eqref{eq:(7)} to Eq. \eqref{eq:b_0k_legacy} leads to
\begin{align}
b_{0,k}^{\text{SLD}} 
%&= \frac{W_0 - k}{W_0} \cdot (1 - \gamma) \cdot \left[ \frac{\gamma}{1 - \gamma} + 1 \right] b_{0,0}^{\text{SLD}} \notag \\
&= \frac{W_0 - k}{W_0} \cdot b_{0,0}^{\text{SLD}} , \quad 1 \leq k \leq W_0 - 1 .
	%\quad \textcolor{red}{\text{... (12)}} 
	\label{eq:(12)}
\end{align}

Next, we try to derive $b_{i,k}^{\text{MLD}}$'s %the MLO part's stationary probabilities 
for $1 \leq i \leq m$. 
First, the state transitions from/to each state with $i=m$ gives
\begin{align}
b_{m,k}^{\text{MLD}} =	\frac{W_m-k}{W_m} \cdot \left[\frac{(p_{\text{AP}}^{\text{MLD}})^m}{1-p_{\text{AP}}^{\text{MLD}}} \cdot b_{0,0}^{\text{MLD}} + b_{m',0}^{\text{MLD}} \right], \notag \\ \quad 1 \leq k \leq W_m-1 . 
	%\quad \textcolor{red}{\text{... (05)}}
	\label{eq:(05)} 
\end{align}
\begin{comment}
\begin{align}
b_{m,k}^{\text{MLD}} &= \frac{W_m-k}{W_m} \bigg[ (1-X) \cdot p_{\text{AP}}^{\text{MLD}} \cdot (p_{\text{AP}}^{\text{MLD}})^{m-1} \cdot b_{0,0}^{\text{MLD}} \notag \\
	&\quad+ \{ 1 - (1-X) \cdot (1-p_{\text{AP}}^{\text{MLD}})\} \cdot \frac{(p_{\text{AP}}^{\text{MLD}})^m}{1-p_{\text{AP}}^{\text{MLD}}} \cdot b_{0,0}^{\text{MLD}} \bigg], \notag \\
&= \frac{W_m-k}{W_m} \cdot \frac{(p_{\text{AP}}^{\text{MLD}})^m}{1-p_{\text{AP}}^{\text{MLD}}} \cdot b_{0,0}^{\text{MLD}}, \quad 1 \leq k \leq W_m-1 . 
	\quad \textcolor{red}{\text{... (05)}}
	\label{eq:(05)} 
\end{align}
\end{comment}
In addition, $b_{m',0}$ can be derived as
\begin{align}
1 \cdot b_{m',0}^{\text{MLD}} &= (\frac{p_{\text{AP}}^{\text{MLD}}}{W_m} \cdot  b_{m-1,0}^{\text{MLD}} + \frac{p_{\text{AP}}^{\text{MLD}}}{W_m} \cdot b_{m,0}^{\text{MLD}} 
 +  b_{m,1}^{\text{MLD}}) \cdot X_\text{AP} \notag \\
 &\quad + \frac{1}{W_m} \cdot b_{m',0}^{\text{MLD}} .
%1 \cdot b_{m',0}^{\text{MLD}} &= \frac{p_{\text{AP}}^{\text{MLD}}}{W_m} \cdot  b_{m-1,0}^{\text{MLD}} + \frac{p_{\text{AP}}^{\text{MLD}}}{W_m} \cdot b_{m,0}^{\text{MLD}} 
%+ X_\text{AP} \cdot b_{m,1}^{\text{MLD}} \notag \\
%&\quad + \frac{1}{W_m} \cdot b_{m',0}^{\text{MLD}} . 
%1 \cdot b_{m',0}^{\text{MLD}} &= \frac{p_{\text{AP}}^{\text{MLD}}}{W_m} b_{m-1,0}^{\text{MLD}} + \frac{p_{\text{AP}}^{\text{MLD}}}{W_m} b_{m,0}^{\text{MLD}} + X b_{m,1}^{\text{MLD}} + \frac{1}{W_m} b_{m',0}^{\text{MLD}} .
%
%1 \cdot b_{m',0} &= \frac{p_{\text{AP}}^{\text{MLD}}}{W_m} \cdot b_{m-1,0}^{\text{MLD}} + \frac{p_{\text{AP}}^{\text{MLD}}}{W_m} \cdot b_{m,0}^{\text{MLD}} + X \cdot b_{m,1}^{\text{MLD}} \notag \\
%&\quad+ \frac{1}{W_m} \cdot b_{m',0} .
\end{align}
Applying Eqs. \eqref{eq:b_MLD_1}, \eqref{eq:b_MLD_2}, and \eqref{eq:(05)} to the above provides
\begin{align}
b_{m',0}^{\text{MLD}} &= \frac{W_m}{W_m - 1} \cdot \frac{X_\text{AP}}{1 - X_\text{AP}} \cdot \frac{(p_{\text{AP}}^{\text{MLD}})^m}{1 - p_{\text{AP}}^{\text{MLD}}} \cdot b_{0,0}^\text{MLD} . 
	\label{eq:b_mprime0}
\end{align}
Then, by combining Eqs. \eqref{eq:(05)} and \eqref{eq:b_mprime0}, we conclude that
\begin{align}
b_{m,k}^{\text{MLD}} &= \frac{W_m - k}{W_m}  \left[ 1 + \frac{X_\text{AP}}{1 - X_\text{AP}} \cdot \frac{W_m}{W_m - 1} \right] 
	\frac{(p_{\text{AP}}^{\text{MLD}})^m}{1 - p_{\text{AP}}^{\text{MLD}}} \; b_{0,0}^{\text{MLD}} ,  \notag \\ 
&\qquad\qquad\qquad\qquad\qquad\qquad 0 \leq k \leq W_m-1.
\end{align}
\begin{comment}
By combining Eqs. \eqref{eq:b_MLD_2} and \eqref{eq:(05)}, we conclude that
\begin{align}
b_{m,k}^{\text{MLD}} &= \frac{W_m-k}{W_m} \cdot \frac{(p_{\text{AP}}^{\text{MLD}})^m}{1-p_{\text{AP}}^{\text{MLD}}} \cdot b_{0,0}^{\text{MLD}}, \quad 0 \leq k \leq W_m-1. \tag{16}
\end{align}
\end{comment}

Second, following the procedure above similarly, the state transitions at each state for $1 \leq i \leq m-1$ provide
\begin{comment}
\textcolor{magenta}{	%%%%%%%%%%%%%%%%%%%%%%%%
\begin{align}
1 \cdot \; b_{i, W_i-1}^{\text{MLD}} & = \frac{p_{\text{AP}}^{\text{MLD}}}{W_i}\cdot b_{i-1,0}^{\text{MLD}} + \frac{1}{W_i} \cdot b_{i',0}^{\text{MLD}} \notag \\ 
&= \frac{1}{W_i} \cdot \left[p_{\text{AP}}^{\text{MLD}} \cdot b_{i-1,0}^\text{MLD} + b_{i',0}^{\text{MLD}} \right] \notag \\
1 \cdot \; b_{i, W_i-2}^{\text{MLD}} & = \frac{p_{\text{AP}}^{\text{MLD}}}{W_i}\cdot b_{i-1,0}^{\text{MLD}} + \frac{1}{W_i} \cdot b_{i',0}^{\text{MLD}} + 1 \cdot b_{i, W_i-1}^{\text{MLD}} \notag \\ 
&= \frac{2}{W_i} \cdot \left[p_{\text{AP}}^{\text{MLD}} \cdot b_{i-1,0}^\text{MLD} + b_{i',0}^{\text{MLD}} \right] \notag \\
&\vdots \notag \\
1 \cdot \; b_{i, 1}^{\text{MLD}} & = \frac{W_i-1}{W_i} \cdot \left[p_{\text{AP}}^{\text{MLD}} \cdot b_{i-1,0}^\text{MLD} + b_{i',0}^{\text{MLD}} \right], \notag
\end{align}
from which we obtain
}	%%%%%%%%%%%%%%%%%%%%%%%%%%%%%%%%%%%%%%%%%%%%
\end{comment}
\begin{align}
b_{i,k}^\text{MLD} = \frac{W_i-k}{W_i} \left[p_{\text{AP}}^{\text{MLD}} \cdot b_{i-1,0}^\text{MLD} + b_{i',0}^{\text{MLD}} \right], \quad 0 \leq k \leq W_i-1.
\end{align}
By combining the above with Eq. \eqref{eq:b_MLD_1}, we get
\begin{align}
b_{i,k}^\text{MLD} = \frac{W_i-k}{W_i} \left[(p_{\text{AP}}^{\text{MLD}})^i \cdot b_{0,0}^\text{MLD} + b_{i',0}^{\text{MLD}} \right], \; 0 \leq k \leq W_i-1 . \label{eq:b_MLD_ik}
\end{align}
Moreover, the state transitions regarding $b_{i',0}^{\text{MLD}}$ lead to
\begin{align}
1 \cdot b_{i',0}^{\text{MLD}} = X_\text{AP} \cdot \frac{p_{\text{AP}}^{\text{MLD}}}{W_i} \cdot b_{i-1,0}^\text{MLD} + X_\text{AP} \cdot b_{i,1}^\text{MLD} + \frac{1}{W_i} \cdot b_{i',0}^{\text{MLD}}.
\end{align}
By applying \eqref{eq:b_MLD_1} and \eqref{eq:b_MLD_ik} to the above, we obtain
\begin{align}
b_{i',0}^{\text{MLD}}  = \frac{W_i}{W_i-1} \cdot \frac{X_\text{AP}}{1-X_\text{AP}} \cdot (p_{\text{AP}}^{\text{MLD}})^i \cdot b_{0,0}^\text{MLD} . \label{eq:b_MLD_iprime0}
\end{align}
\begin{comment}
\begin{align}
1 \cdot \; b_{i, W_i-1}^{\text{MLD}} &= \frac{1}{W_i} 
	\left[ (1-X) \cdot p_{\text{AP}}^{\text{MLD}} \cdot b_{i-1,0}^{\text{MLD}} \right. \notag \\ 
	&\quad+ \left. X \cdot (1-p_{\text{AP}}^{\text{MLD}}) \cdot b_{i,0}^{\text{MLD}} \right], \notag \\
1 \cdot \; b_{i, W_i-2}^{\text{MLD}} &= \frac{1}{W_i} \cdot \left[ (1-X) \cdot p_{\text{AP}}^{\text{MLD}} \cdot b_{i-1,0}^{\text{MLD}} \right. \notag \\ 
	&\quad+ \left. X \cdot (1-p_{\text{AP}}^{\text{MLD}}) \cdot b_{i,0}^{\text{MLD}} \right] 
		+ \textcolor{red}{\frac{1}{W_i}} \cdot b_{i, W_i-1}^{\text{MLD}} \notag \\
	&= \frac{2}{W_i} \cdot \left[ (1-X) \cdot p_{\text{AP}}^{\text{MLD}} \cdot b_{i-1,0}^{\text{MLD}} \right. \notag \\ 
	&\quad+ \left. X \cdot (1-p_{\text{AP}}^{\text{MLD}}) \cdot b_{i,0}^{\text{MLD}} \right], \notag \\
&\vdots \notag \\
1 \cdot \; b_{i,1}^{\text{MLD}} &= \frac{W_i-1}{W_i} \cdot \left[ (1-X) \cdot p_{\text{AP}}^{\text{MLD}} \cdot b_{i-1,0}^{\text{MLD}} \right. \notag \\ 
	&\quad+ \left. X \cdot (1-p_{\text{AP}}^{\text{MLD}}) \cdot b_{i,0}^{\text{MLD}} \right], \notag
\end{align}
\end{comment}
Finally, applying Eq. \eqref{eq:b_MLD_iprime0} to Eq. \eqref{eq:b_MLD_ik} gives us 
\begin{align}
b_{i,k}^{\text{MLD}} = \frac{W_i-k}{W_i} \cdot \left[ 1 + \frac{X_\text{AP}}{1-X_\text{AP}} \cdot \frac{W_i}{W_i-1}\right] \cdot (p_{\text{AP}}^{\text{MLD}})^i
 \cdot b_{0,0}^\text{MLD}, \notag \\
  \quad 1 \leq k \leq W_i-1.
\end{align}
\begin{comment}
\begin{align}
b_{i,k}^{\text{MLD}} &= \frac{W_i-k}{W_i} 
	\left[ (1-X) \cdot (p_{\text{AP}}^{\text{MLD}})^i \right. \notag \\
	&\quad+ \left. X \cdot (1-p_{\text{AP}}^{\text{MLD}}) \right] b_{0,0}^{\text{MLD}}, \notag \\
	&= \frac{W_i-k}{W_i} \cdot (p_{\text{AP}}^{\text{MLD}})^i \cdot b_{0,0}^{\text{MLD}}, \quad 1 \leq k \leq W_i-1. 
	\quad \textcolor{red}{\text{... (07)}}
	\label{eq:(07)} 
\end{align}
\end{comment}
\begin{comment}
By combining Eqs. \eqref{eq:(07)} and \eqref{eq:b_MLD_1}, we finally obtain
\begin{align}
b_{i,k}^{\text{MLD}} = \frac{W_i-k}{W_i} \cdot (p_{\text{AP}}^{\text{MLD}})^i \cdot b_{0,0}^\text{MLD}, 
	\quad 
	&1 \leq i \leq m-1, \notag \\ 
	&0 \leq k \leq W_i-1 . 
		\quad \textcolor{red}{\text{... (08)}}
		\label{eq:(08)} 
\end{align}
\end{comment}

In the SLO part, for $1 \leq i \leq m$, the corresponding MC closely follows the MC structure in \cite{bianchi2000performance}, from which we can obtain 
\begin{align}
	b_{i,k}^{\text{SLD}} = \frac{W_i - k}{W_i} \cdot b_{i,0}^{\text{SLD}} , 
	\quad 1 \leq i \leq m , \quad 0 \leq k \leq W_i-1 .
\end{align}
where
\begin{align}
b_{i,0}^{\text{SLD}} &= (p_{\text{AP}}^{\text{SLD}})^i \cdot b_{0,0}^{\text{SLD}}, \quad 1 \leq i \leq m-1, \notag \\
b_{m,0}^{\text{SLD}} &= \frac{(p_{\text{AP}}^{\text{SLD}})^m}{1-p_{\text{AP}}^{\text{SLD}}} \cdot b_{0,0}^{\text{SLD}}, \quad i = m . 
\label{eq:b_SLO_i0_m0}
\end{align}
Since the sum of stationary probabilities should be 1, we get
\begin{align}
1 &= \sum_{i=0}^{m} \sum_{k=0}^{W_i - 1} b_{i,k}^{\text{MLD}} 
	+ \sum_{i'=0'}^{m'} b_{i',0}^{\text{MLD}} 
	+ \sum_{i=0}^{m} \sum_{k=0}^{W_i - 1} b_{i,k}^{\text{SLD}} \notag \\
&= \sum_{k=0}^{W_0 - 1} b_{0,k}^{\text{MLD}} 
	+ \sum_{k=0}^{W_m - 1} b_{m,k}^{\text{MLD}} 
	+ \sum_{i=1}^{m-1} \sum_{k=0}^{W_i - 1} b_{i,k}^{\text{MLD}} 
	+ \sum_{i'=0'}^{m'} b_{i',0}^{\text{MLD}}  \notag \\
	&\quad+ \sum_{k=0}^{W_0 - 1} b_{0,k}^{\text{SLD}} 
	+ \sum_{k=0}^{W_m - 1} b_{m,k}^{\text{SLD}} 
	+ \sum_{i=1}^{m-1} \sum_{k=0}^{W_i - 1} b_{i,k}^{\text{SLD}} .
\end{align}
By applying Eqs. \eqref{eq:(9)} through \eqref{eq:b_SLO_i0_m0}
%\textcolor{green}{Eqs. (27), (28), (30), (34), (35), (39), (40), (41) and (42) } 
to the above, we obtain
\begin{align}
1 &= \textcolor{black}{\lambda_1} \cdot b_{0,0}^{\text{MLD}} + \textcolor{black}{\lambda_2} \cdot b_{0,0}^{\text{SLD}} , \\
\textcolor{black}{\lambda_1} &:= \left[ \frac{1}{1 - p_{\text{AP}}^{\text{MLD}}} + \sum_{i=0}^{m-1} \left( \frac{W_i - 1}{2} + \frac{X_\text{AP}}{1 - X_\text{AP}} \frac{W_i}{2} \right) (p_{\text{AP}}^{\text{MLD}})^i \right. \notag \\
&\quad +  \frac{W_m - 1}{2} + \frac{X_\text{AP}}{1 - X_\text{AP}} \left( \frac{W_m}{2} \cdot \frac{(p_{\text{AP}}^{\text{MLD}})^m}{1 - p_{\text{AP}}^{\text{MLD}}} \right. \notag \\
&\quad + \left. \left.  \sum_{i=0}^{m-1}\frac{W_i \cdot (p_{\text{AP}}^{\text{MLD}})^i}{W_i -1} + \frac{W_m}{W_m - 1} \cdot \frac{(p_{\text{AP}}^{\text{MLD}})^m}{1 - p_{\text{AP}}^{\text{MLD}}}\right) \right],  \\
%\textcolor{black}{\lambda_1} &:= \left[ \frac{1}{1 - p_{\text{AP}}^{\text{MLD}}} + \frac{W_0 - 1}{2} + \frac{X_\text{AP}}{1 - X_\text{AP}} \cdot \frac{W_0}{2} \right. \notag \\
%	&\quad + \sum_{i=1}^{m-1} \left( \frac{W_i - 1}{2} + \frac{X_\text{AP}}{1 - X_\text{AP}} \cdot \frac{W_i}{2} \right) \cdot (p_{\text{AP}}^{\text{MLD}})^i \notag \\
%	&\quad + \left. \frac{W_m - 1}{2} + \frac{X_\text{AP}}{1 - X_\text{AP}} \cdot \frac{W_m}{2} \cdot \frac{(p_{\text{AP}}^{\text{MLD}})^m}{1 - p_{\text{AP}}^{\text{MLD}}} \right], \\
%
\textcolor{black}{\lambda_2} &:= \bigg[ \frac{W_0 + 1}{2} + \frac{W_m + 1}{2} \frac{(p_{\text{AP}}^{\text{SLD}})^m}{1 - p_{\text{AP}}^{\text{SLD}}} 
	+ \sum_{i=1}^{m-1} \frac{W_i + 1}{2} (p_{\text{AP}}^{\text{SLD}})^i \bigg] . 
\end{align}
Then, applying Eq. \eqref{eq:(7)} to the above leads to 
\begin{comment}
\begin{align}
%1 &= \textcolor{black}{\lambda_1} \cdot b_{0,0}^{\text{MLD}} + \textcolor{black}{\lambda_2} \cdot b_{0,0}^{\text{SLD}} \\
%&\Rightarrow 
1 = \left[ \textcolor{black}{\lambda_1} \cdot \frac{1-\gamma}{\gamma} + \textcolor{black}{\lambda_2} \right] \cdot b_{0,0}^{\text{SLD}} \text{ or } %\notag \\
%&\quad \;\; 
1 = \left[ \textcolor{black}{\lambda_1} + \textcolor{black}{\lambda_2} \cdot \frac{\gamma}{1-\gamma} \right] \cdot b_{0,0}^{\text{MLD}} .
\end{align}
\end{comment}
\begin{align}
b_{0,0}^{\text{SLD}} &= \left[ \textcolor{black}{\lambda_1} \frac{\gamma}{1 - \gamma} + \textcolor{black}{\lambda_2} \right]^{-1} , \; %\\
b_{0,0}^{\text{MLD}} &= \left[ \textcolor{black}{\lambda_1} + \frac{1 - \gamma}{\gamma} \textcolor{black}{\lambda_2} \right]^{-1} . \notag
%b_{0,0}^{\text{SLD}} &= \frac{1}{\textcolor{black}{\lambda_1} \cdot \frac{\gamma}{1 - \gamma} + \textcolor{black}{\lambda_2}} \\
%b_{0,0}^{\text{MLD}} &= \frac{1}{\textcolor{black}{\lambda_1} + \frac{1 - \gamma}{\gamma} \cdot \textcolor{black}{\lambda_2}} .
\end{align}

%--------------------------------------------------------------
\subsection{Derivation of transmit probabilities}

We are now ready to derive the transmit probability of an AP MLD towards a Non-AP MLD STA and a legacy STA, denoted by $\tau_{\text{AP}}^{\text{MLD}}$ and $\tau_{\text{AP}}^{\text{SLD}}$ respectively, such as
\begin{align}
\tau_{\text{AP}}^{\text{MLD}} 
	&= \sum_{i=0}^{m} b_{i,0}^{\text{MLD}} 
	= \frac{b_{0,0}^{\text{MLD}}}{1 - p_{\text{AP}}^{\text{MLD}}} 
	= \frac{
		{1}/{(1 - p_{\text{AP}}^{\text{MLD}})}
		}{
		\textcolor{black}{\lambda_1} + \frac{1 - \gamma}{\gamma} \cdot \textcolor{black}{\lambda_2}
		} , 
	\label{eq:tau_AP_MLD} \\
\tau_{\text{AP}}^{\text{SLD}} 
	&= \sum_{i=0}^{m} b_{i,0}^{\text{SLD}} 
	= \frac{b_{0,0}^{\text{SLD}}}{1 - p_{\text{AP}}^{\text{SLD}}} 
	= \frac{
		{1}/{(1 - p_{\text{AP}}^{\text{SLD}})}
		}{
		\textcolor{black}{\lambda_1} + \frac{\gamma}{1 - \gamma} \cdot \textcolor{black}{\lambda_2}
		} .
	\label{eq:tau_AP_legacy}
\end{align}

%% file: nonap_mc.tex
\begin{figure}
%	\centerline{\includegraphics[width=0.8\linewidth]{figures/Fig5_0_correct_backoff_operation_UL.eps}}
	\centerline{\includegraphics[width=0.8\linewidth]{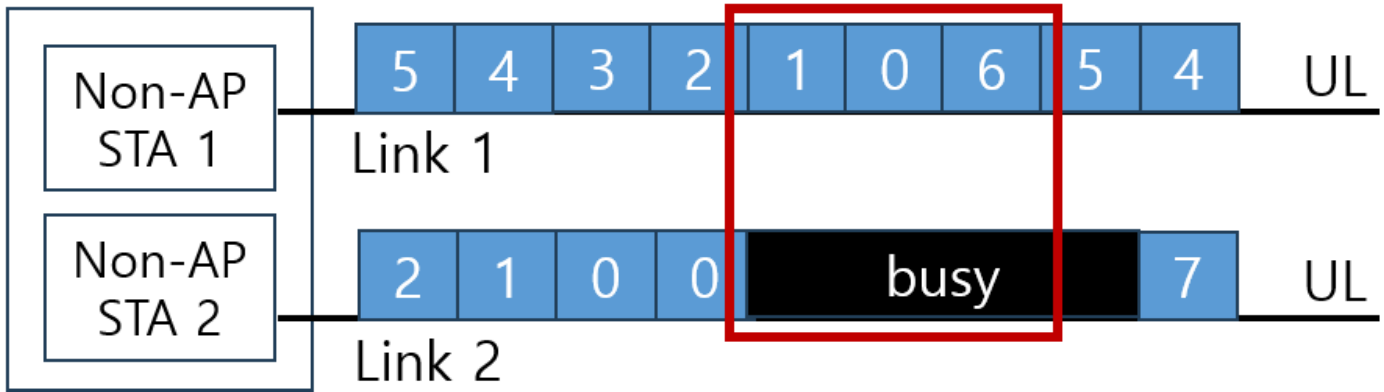}}
	\caption{Standard-compliant backoff operation during UL transmission}
	\label{fig:correct_UL}
	\vspace{-0.1in}
\end{figure}

\begin{figure}
%	\centerline{\includegraphics[width=0.99\linewidth]{figures/Fig7_distinction_state_0_MLD.eps}}
	\centerline{\includegraphics[width=0.99\linewidth]{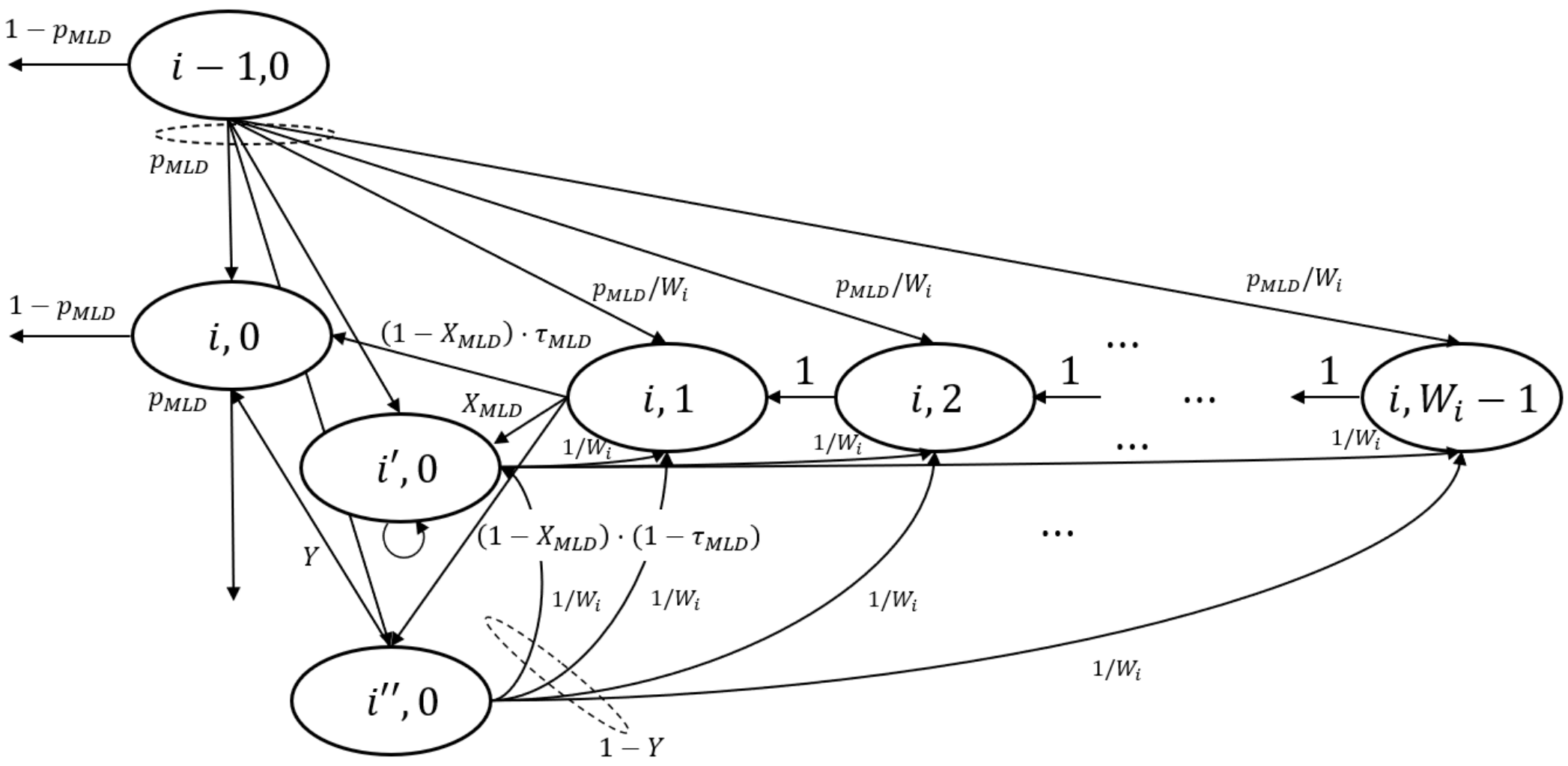}}
	\caption{Separation of state $(i,0)$ into $(i,0)$, $(i',0)$ and $(i'',0)$}
	\label{fig:distinction_MLD}
	\vspace{-0.1in}
\end{figure}

\begin{figure}
%	\centerline{\includegraphics[width=0.99\linewidth]{figures/Fig8_MC_for_non_AP_MLD.eps}}
	\centerline{\includegraphics[width=0.99\linewidth]{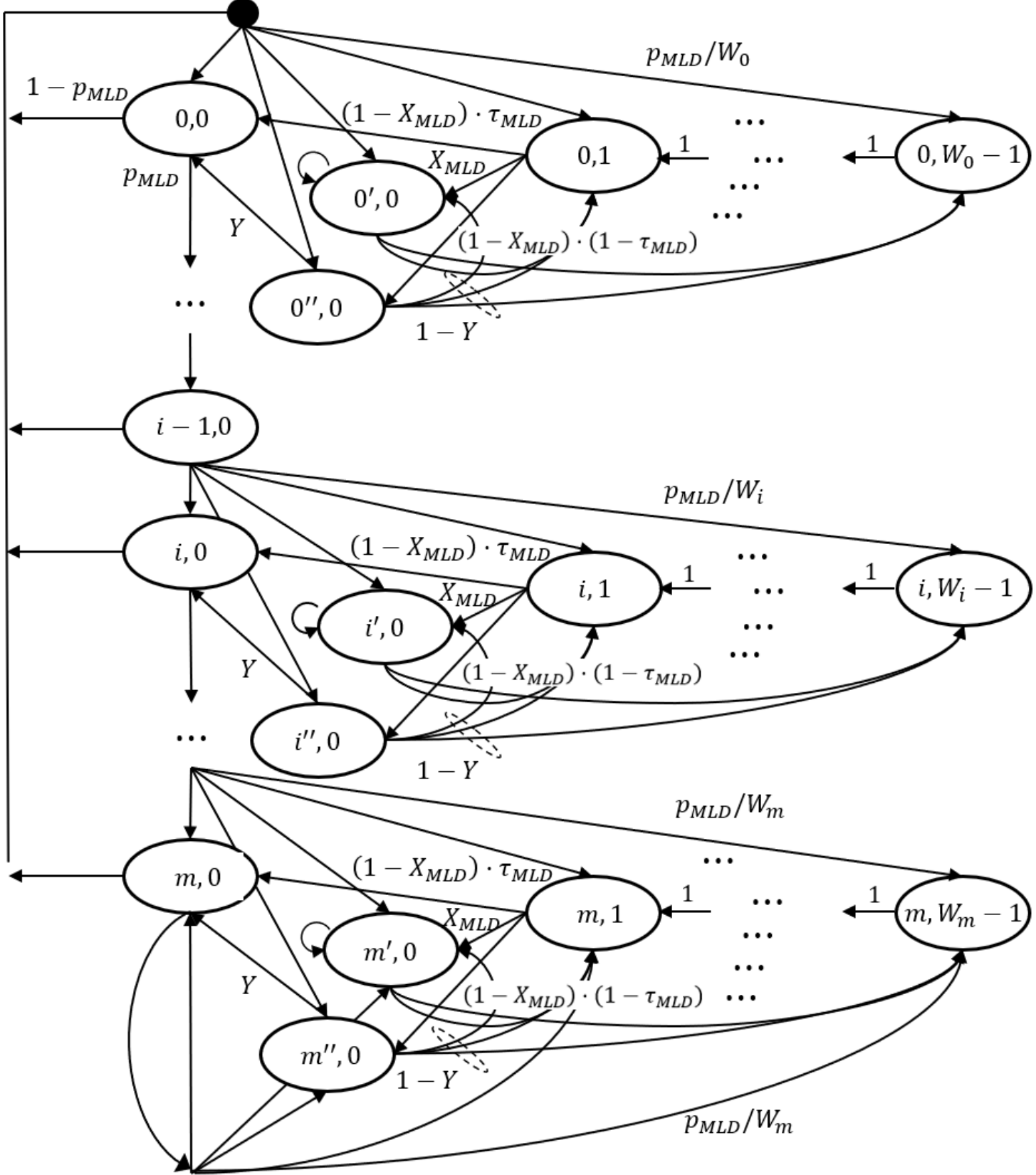}}
	\caption{The proposed MC for the non-AP MLD}
	\label{fig:non_AP_MLD}
	\vspace{-0.1in}
\end{figure}

%First, we explain the differences in start-time alignment for non-AP MLDs between the Wi-Fi 7 standard and the model proposed in \cite{song2021performance}. 
To better explain our proposed MC in this section, suppose a non-AP MLD's backoff counter for Link 2 reaches zero first and then it is waiting until Link 1's backoff counter also reaches zero. 
While Link 2 is waiting, it becomes busy and later on Link 1's backoff counter reaches zero, as depicted in Fig. \ref{fig:correct_UL}.
In such a case, the standard mandates that the backoff process should restart on Link 1 maintaining the same backoff stage. 
Nevertheless, the start-time alignment model proposed in \cite{song2021performance} failed to account for such possibility, leading to inaccuracy under saturated conditions due to frequent channel state changes.

To address this issue in the non-AP MLD MC, we introduce a new state $(i',0)$ and a state transition from $(i,1)$ to $(i',0)$, %similar to the AP MLD's MC proposed in the previous section. 
%This transition allows 
to allow the backoff process to restart when the backoff counter reaches zero, provided that the other channel busy. %deviating from the original Bianchi model.
Additionally, to accurately model start-time alignment, we introduce another state $(i'',0)$ and a state transition from $(i,1)$ to $(i'',0)$ for the situation where one link's backoff counter reaches zero while the other link's backoff counter has not yet done so.
In this case, the link with a zero backoff counter enters $(i'',0)$ and stays there as long as both links are idle.
If either link becomes busy during this waiting period, however, the backoff restarts while retaining the current backoff stage (by making a transition from $(i'',0)$ to either $(i',0 )$, $(i,1)$, $\ldots$, or $(i,W_{i}-1)$); if both links remain idle until the other link's backoff counter also reaches zero, the transmission proceeds (by making a transition from $(i'',0)$ to $(i,0)$).

Fig. \ref{fig:distinction_MLD} illustrates the aforementioned mechanism.
%There are three possible state transitions from $(i,1)$ to either $(i,0)$, $(i',0)$ or $(i'',0)$, each corresponding to transmission, restarted backoff, and waiting for the other link, respectively. 
%
In the figure, $X_{\text{MLD}}$ represents the probability that the other link is busy by other devices and $Y$ denotes the probability that a specific link's backoff counter remains at zero until the other link's backoff counter reaches zero.
%As with the AP MLD, we construct individual MCs for each STA within the non-AP MLD, ensuring that each STA performs its backoff process based on its link state.
Building upon the above, the structure of our proposed MC for each non-AP MLD is illustrated in Fig. \ref{fig:non_AP_MLD}.

\subsection{Derivation of stationary probabilities} %for the non-AP MLD.}\\

In this section, we derive the stationary probabilities of the non-AP MLD's MC in Fig. \ref{fig:non_AP_MLD}, which are denoted by $b_{i,k}$.
After deriving $b_{i,k}$'s, we will present how to derive the two transmit probabilities $\tau_{\text{MLD}}^{I}$ and $\tau_{\text{MLD}}^{II}$, using $p_{\text{MLD}}^I$ and $p_{\text{MLD}}^{II}$ respectively.
Note that earlier in Section \ref{sec:ap_mc} we distinguished Case 1 from Case 2 according to the packet destination, with two types of transmission probabilities $\tau_{\text{AP}}^{\text{SLD}}$ and $\tau_{\text{AP}}^\text{MLD}$. 
The structure of the non-AP MLD's MC, however, remains unchanged whether the system operates under Case 1 or Case 2, and thus the derivation steps in the sequel would be identical for both cases.
In such a vein, we will use the notations $\tau_{\text{MLD}}$ and $p_{\text{MLD}}$ without specifying the corresponding case with a superscript like ${I}$ or ${II}$.\footnote{It should be emphasized that $p_{\text{MLD}}^{I}$ should be used when deriving $\tau_{\text{MLD}}^{I}$, whereas $p_{\text{MLD}}^{II}$ should be used when deriving $\tau_{\text{MLD}}^{II}$.}

First, the vertical state transitions among the leftmost states $(i,0)$, $0 \le i \le m$ provide
\begin{align}
b_{i-1,0} \cdot p_{\text{MLD}} &= b_{i,0} , \quad 0 < i < m , \notag \\
	&\Rightarrow b_{i,0} = p_{\text{MLD}}^i \cdot b_{0,0} , \quad 0 \leq i < m ,
	%\quad \textcolor{red}{\cdots (1)} 
	\label{eq:IV(1)} \\
b_{m-1,0} \cdot p_{\text{MLD}} &= (1 - p_{\text{MLD}}) \cdot b_{m,0} \notag \\
	&\Rightarrow b_{m,0} = \frac{(p_{\text{MLD}})^m}{1 - p_{\text{MLD}}} \cdot b_{0,0} , 
	%\quad \textcolor{red}{\cdots (2)} 
	\label{eq:IV(2)}
\end{align}
By Eqs. \eqref{eq:IV(1)} and \eqref{eq:IV(2)}, we obtain
\begin{align}
\sum_{i=0}^{m} b_{i,0} 
%	&= \sum_{i=0}^{m-1} b_{i,0} + \frac{p^m}{1 - p} \cdot b_{0,0} \notag \\
	&= \frac{1 - (p_{\text{MLD}})^m}{1 - p_{\text{MLD}}} \cdot b_{0,0} + \frac{(p_{\text{MLD}})^m}{1 - p_{\text{MLD}}} \cdot b_{0,0} %\notag \\
	%&= 
	= \frac{b_{0,0}}{1 - p_{\text{MLD}}} . 
	%\quad \textcolor{red}{\cdots (3)} 
	\label{eq:IV(3)}
\end{align}

\textcolor{black}{Recall from Section \ref{sec:model} that a non-AP MLD must align both start and end times to avoid IDC interference.
When one link's backoff counter reaches zero, it holds until the other link also reaches zero.
If a busy channel is sensed during this period, the link with a zero backoff counter restarts its backoff.
To incorporate this behavior into the MC model, we define two probabilities $X_{\text{MLD}}$ and $Y$.}
First, $X_{\text{MLD}}$ is `the probability that from the viewpoint of a certain link belonging to a non-AP MLD, the other link is busy due to the transmission either from AP MLD, legacy STA or other non-AP MLDs'.
In addition, $Y$ represents `the probability that, within a non-AP MLD, a specific link's backoff counter remains at zero until the other link's backoff counter reaches zero as well'.\footnote{Similar to $X_{\text{AP}}$, the derivations of  $X_{\text{MLD}}$ and $Y$ require $\tau$'s and $p$'s and thus will be presented later in Section \ref{subsec:derivation_x_y}.} 
Although $X_{\text{MLD}}$ remains unchanged regardless of Case 1 or Case 2, $Y$ could vary, and hence %ies depending on the case of the AP MLD. 
%Although $X_{\text{MLD}}$ remains unchanged regardless of whether the AP MLD is in Case 1 or Case 2, $Y$ could vary. %ies depending on the case of the AP MLD. 
$Y_{{case}_1}$ must be used when deriving $\tau_{\text{MLD}}^{I}$ whereas $Y_{{case}_2}$ must be used when deriving $\tau_{\text{MLD}}^{II}$.
Nevertheless, due to the same reason previously explained for $\tau_{\text{MLD}}$ and $p_{\text{MLD}}$, we will use the notation $Y$ without explicitly differentiating between $Y_{{case}_1}$ and $Y_{{case}_2}$ in the sequel.

For stage 0 (i.e., $i = 0$), \textcolor{black}{transitions to/from each state gives} %\textcolor{magenta}{we obtain from the state transitions to/from each state.
\begin{align}
b_{0,k} &= \frac{W_0 - k}{W_0} \{ b_{0,0} + b_{0^{\prime},0} 
	+ (1 - Y) b_{0^{\prime \prime},0} \} , \notag\\
	&\qquad\qquad\qquad\qquad\qquad\qquad 1 \leq k \leq W_0 - 1 ,
	%\quad \textcolor{red}{\cdots (4)} 
	\label{eq:IV(4)}
\end{align}
\textcolor{black}{by applying g.g.b.e. to each $b_{0,k}$.}
\begin{comment}
\begin{align}
1 \cdot b_{0,W_0 - 1} 
	&= \frac{1}{W_0} \cdot (1 - p_{\text{MLD}}) \cdot \sum_{i=0}^{m} b_{i,0} + \frac{1}{W_0} \cdot b_{0',0} \notag \\ 
	&\quad+ \frac{1}{W_0} \cdot (1 - Y) \cdot b_{0^{\prime \prime},0} \notag \\
	&= \frac{1}{W_0} \cdot \{ b_{0,0} + b_{0',0} + (1 - Y) \cdot b_{0^{\prime \prime},0} \} , \notag \\
1 \cdot b_{0,W_0 - 2} &= \frac{2}{W_0} \cdot \{ b_{0,0} + b_{0^{\prime},0} + (1 - Y) \cdot b_{0^{\prime \prime},0} \} , \notag \\
&\vdots \notag \\
1 \cdot b_{0,1} &= \frac{W_0 - 1}{W_0} \cdot \{ b_{0,0} + b_{0^{\prime},0} + (1 - Y) \cdot b_{0^{\prime \prime},0} \} , \notag
\end{align}
from which we obtain %\textcolor{red}{[$Y$ should have been defined earlier.]} 
\end{comment}

Moreover, the state transitions to/from the state $(0',0)$ gives
\begin{align}
b_{0',0} 
	&= \frac{1}{W_0} \cdot X_{\text{MLD}} \cdot (1 - p_{\text{MLD}}) \cdot \sum_{i=0}^{m} b_{i,0} 
		+ \frac{1}{W_0} \cdot b_{0',0} \notag \\
	&\quad+ \frac{1}{W_0} (1 - Y) \cdot b_{0^{\prime \prime},0} %\notag \\
		+ X_{\text{MLD}} \cdot b_{0,1} , \label{eq:IV(4_1)} %\\
\end{align}
to which we apply Eq. \eqref{eq:IV(4)} to obtain
\begin{align}
b_{0^{\prime},0} &= 
	\bigg( \frac{1}{W_0 - 1} + X_{\text{MLD}} \bigg) 
		\cdot \frac{1 - Y}{1 - X_{\text{MLD}}} 
		\cdot b_{0^{\prime \prime},0} \notag\\
	&\quad+ \frac{X_{MLD}}{1 - X_{\text{MLD}}} \cdot \frac{W_0}{W_0 - 1} \cdot b_{0,0} .	 
	\label{eq:IV(5)}
\end{align}
\begin{comment}
\begin{align}
%\frac{W_0 - 1}{W_0} \cdot b_{0,0} 
%	&= \frac{1}{W_0} X_{MLD} \cdot b_{0,0} \notag \\
%	&\quad+ \frac{W_0 - 1}{W_0} X_{MLD} \{ b_{0,0} 
%		+ b_{0^{\prime},0} \notag \\
%	&\qquad\qquad\qquad\quad+ (1 - Y) \cdot b_{0^{\prime \prime},0} \} , \notag \\
(1 - X_{MLD}) \frac{W_0 - 1}{W_0} \cdot b_{0,0} 
	&= X_{MLD} \cdot b_{0,0} 
		+ X_{MLD} (1 - Y) \cdot b_{0^{\prime \prime},0} , 
	\quad \textcolor{red}{\cdots (5)} \label{eq:IV(5)}
\end{align}
\end{comment}
%\textcolor{red}{[HK: in the above, are the last two equations reduced versions of the first equation?]}

Furthermore, the state transitions to/from the state $(0'',0)$ provides
\begin{align}
b_{0^{\prime \prime},0} &= (1 - X_{\text{MLD}}) \cdot (1 - \tau_{\text{MLD}}) \cdot \frac{1}{W_0} \cdot (1 - p_{\text{MLD}}) \cdot \sum_{i=0}^{m} b_{i,0} \notag \\
	&\quad+ (1 - X_{\text{MLD}}) \cdot (1 - \tau_{\text{MLD}}) \cdot b_{0,1} , %\notag \\
%b_{0^{\prime \prime},0} &= \frac{1}{W_0} \cdot (1 - X_{MLD}) \cdot (1 - T) \cdot b_{0,0} \notag \\
%	&\quad+ \frac{W_0 - 1}{W_0} \cdot (1 - X_{MLD}) \cdot (1 - T) \{ b_{0,0} + b_{0^{\prime},0} \notag \\
%	&\qquad\qquad\qquad\qquad\qquad\qquad\qquad+ (1 - Y) \cdot b_{0^{\prime \prime},0} \} , \notag
\end{align}
%\begin{align}	
%\bigg[ 1 - \frac{W_0 - 1}{W_0} (1 &- X_{MLD}) (1 - T) (1 - Y) \bigg] \cdot b_{0^{\prime \prime},0} \notag \\
%&= (1 - X_{MLD}) (1 - T) \cdot \bigg[ b_{0,0} 
%	+\frac{W_0 - 1}{W_0} \cdot b_{0^{\prime},0} \bigg] , \notag \\
%%&= (1 - X_{MLD}) (1 - T) \cdot b_{0,0} +\frac{W_0 - 1}{W_0} (1 - X_{MLD}) (1 - T) \cdot b_{0^{\prime},0} , \notag
%\end{align}
%\textcolor{red}{[HK: in the above, what is the relationship among the three equations?]}
to which we can apply Eq. \eqref{eq:IV(4)} to obtain
%\begin{align}
%b_{0^{\prime \prime},0} &= \frac{(1 - X_{MLD}) (1 - T)}{1 - \frac{W_0 - 1}{W_0} (1 - X_{MLD}) (1 - T) (1 - Y)} \notag \\
%	&\qquad\cdot \left\{ b_{0,0} + X_{MLD} \cdot b_{0,0} + \frac{1}{W_0} (1 - Y) \cdot b_{0^{\prime \prime},0} \right\} . \notag
%\end{align}
%\begin{align}
%\bigg\{ 1 &- \frac{W_0-1}{W_0} (1-X_{MLD}) (1-\tau) (1-Y) - \frac{1}{W_0} (1-\tau) (1-Y) \notag \\
%	&- \frac{W_0-1}{W_0} X_{MLD} (1-\tau) (1-Y) \bigg\} \cdot b_{0^{\prime \prime},0} 
%	= (1-\tau) \cdot b_{0,0} , \\
%\bigg\{ 1 &- \frac{W_0-1}{W_0} (1-\tau) (1-Y)-\frac{1}{W_0} (1-\tau)(1-Y) \bigg\} \cdot b_{0^{\prime \prime},0} \notag \\
%	&\qquad\qquad\qquad\qquad\qquad\qquad\qquad = (1-\tau) \cdot b_{0,0} , \\
%\{ 1 &- (1-\tau) (1-Y) \} \cdot b_{0^{\prime \prime},0} = (1-\tau) \cdot b_{0,0} .
%\end{align}
%\textcolor{red}{[HK: in the above, what is the relationship among the four equations?]}
%Therefore, we obtain
\begin{align}
b_{0^{\prime \prime},0} = \frac{(1-\tau_{\text{MLD}})}{\{1-(1-\tau_{\text{MLD}}) (1-Y)\}} \cdot b_{0,0} . 
	%\quad \textcolor{red}{\cdots (6)} 
	\label{eq:IV(6)}
\end{align}
\begin{comment}
\textcolor{red}{[HK: how do we obtain the equation below?]}
\begin{align}
b_{0^{\prime},0} &= \frac{X_{MLD}}{1 - X_{MLD}} \cdot \frac{W_0}{W_0 - 1} \cdot b_{0,0} 
	+ \frac{1}{W_0 - 1} \cdot \frac{1 - Y}{1 - X_{MLD}} \cdot b_{0^{\prime\prime},0} \notag \\
	&\quad + \frac{X_{MLD}}{1 - X_{MLD}} \cdot (1 - Y) \cdot b_{0^{\prime \prime},0} \notag \\
&= \left[ \frac{X_{MLD}}{1 - X_{MLD}} \cdot \frac{W_0}{W_0 - 1} + \frac{1 - Y}{W_0 - 1} \cdot \frac{1}{1 - X_{MLD}} \cdot (1 - T) \right. \notag \\
	&\quad \left. + \frac{X_{MLD}}{1 - X_{MLD}} \cdot (1 - Y) \cdot \frac{(1 - T)}{1 - (1 - T) (1 - Y)} \right] \cdot b_{0,0} . 
	\quad \textcolor{red}{\text{... (7)}} \label{eq:IV(7)}
\end{align}
\end{comment}
Then, by applying Eq. \eqref{eq:IV(6)} to \eqref{eq:IV(5)} leads to
\begin{align}
b_{0',0} &= \bigg[ \frac{X_{\text{MLD}}}{1 - X_{\text{MLD}}} \cdot \frac{W_0}{W_0 - 1} + \left(\frac{1}{W_0 - 1}  + X_{\text{MLD}}\right)\lambda_3 \bigg] b_{0,0} ,\label{eq:IV(7)}
%b_{0',0} &= \bigg[ \frac{X_{\text{MLD}}}{1 - X_{\text{MLD}}} \cdot \frac{W_0}{W_0 - 1} + \frac{1 - Y}{1 - X_{\text{MLD}}} \cdot \frac{\lambda_3}{W_0 - 1}  \notag\\
%	&\quad+ X_{\text{MLD}} \cdot \textcolor{black}{\lambda_3}	
%	\bigg] \cdot b_{0,0} ,
%	\label{eq:IV(7)}
%b_{0',0} &= \left[ \frac{X_{MLD}}{1 - X_{MLD}} \cdot \frac{W_0}{W_0 - 1} + \frac{1 - Y}{W_0 - 1} \cdot \frac{1}{1 - X_{MLD}} \cdot (1 - \tau) \right. \notag \\
%	&\quad \left. + \frac{X_{MLD}}{1 - X_{MLD}} \cdot (1 - Y) \cdot \frac{(1 - \tau)}{1 - (1 - \tau) (1 - Y)} \right] \cdot b_{0,0} . \label{eq:IV(7)}
\end{align}
where
\begin{align}
\textcolor{black}{\lambda_3} := \frac{1-\tau_{\text{MLD}}}{1-(1-\tau_{\text{MLD}})(1-Y)} 
	\cdot \frac{1-Y}{1-X_{\text{MLD}}} . %\notag
\end{align}
By combining Eqs. \eqref{eq:IV(4)}, \eqref{eq:IV(6)} and \eqref{eq:IV(7)}, we obtain %(4), (6) and (7),
\begin{comment}
\begin{align}
b_{0,k} 
%&= \frac{W_0-k}{W_0} \cdot \bigg[ 1  +\frac{X_{MLD}}{1-X_{MLD}} \cdot \frac{W_0}{W_0-1} \notag \\
%	&\qquad\qquad+\frac{1}{W_0-1} \cdot \frac{1-Y}{1-X_{MLD}} \cdot \frac{1-\tau}{1-(1-\tau) (1-Y)} \notag \\
%	&\qquad\qquad+\frac{X_{MLD}}{1-X_{MLD}} \cdot(1-Y) \cdot \frac{1-\tau}{1-(1-\tau) (1-Y)} \notag \\
%	&\qquad\qquad+ \frac{(1-Y)}{1-(1-\tau) (1-Y)} \bigg] \cdot b_{0,0} \notag \\
&= \frac{W_0-k}{W_0} \cdot \bigg[ 1 + \frac{X_{MLD}}{1-X_{MLD}} \cdot \frac{W_0}{W_0-1} \notag \\
	&\quad+ \frac{1-Y}{1-X_{MLD}} \cdot \frac{W_0}{W_0-1} \cdot 
		\frac{1-\tau}{1-(1-\tau) (1-Y)} \bigg] \cdot b_{0.0}
\end{align}
For $1 \leq k \leq W_0-1$, we obtain
\end{comment}
\begin{align}
b_{0,k} &= \frac{W_0-k}{W_0} \cdot \bigg[ 
	1 + \frac{W_0}{W_0-1} \bigg( \frac{X_{\text{MLD}}}{1-X_{\text{MLD}}} + \textcolor{black}{\lambda_3} \bigg)
		\bigg] \cdot b_{0.0} , \notag \\
	&\qquad\qquad\qquad\qquad\qquad\qquad\quad 1 \leq k \leq W_0-1 . 
	%\textcolor{red}{\cdots (8)} 
	\label{eq:IV(8)}
%b_{0,k} &= \frac{W_0-k}{W_0} \cdot \left[ 
%	1 + \frac{X_{MLD}}{1-X_{MLD}} \cdot \frac{W_0}{W_0-1} \right. \notag \\
%	&\quad+ \left. \frac{1-Y}{1-X_{MLD}} \cdot \frac{W_0}{W_0-1} \cdot \frac{1-\tau}{1 - (1-\tau) (1-Y)}
%		\right] \cdot b_{0.0} , \notag \\
%	&\qquad\qquad\qquad\qquad\qquad\qquad\qquad 1 \leq k \leq W_0-1 . 
%	\textcolor{red}{\cdots (8)} \label{eq:IV(8)}
\end{align}

Next, for stage $m$ (i.e., $i = m$), the state transitions \textcolor{black}{yield}
%to/from each state gives
\begin{comment}
\begin{align}
1 &\cdot b_{m, W_m - 1} 
	= \frac{p_{\text{MLD}}}{W_m} \cdot b_{m - 1, 0} + \frac{1}{W_m} \cdot b_{m', 0} \notag \\ 
	&\qquad\qquad\qquad+ \frac{1}{W_m} \cdot (1 - Y) \cdot b_{m'', 0} 
		+ \frac{p_{\text{MLD}}}{W_m} \cdot b_{0, 0} \notag \\
	&\qquad\qquad= \frac{1}{W_m} \left[ \frac{(p_{\text{MLD}})^m}{1 - p_{\text{MLD}}} b_{0, 0} + b_{m', 0} + (1 - Y) b_{m'', 0} \right] , \notag \\
1 &\cdot b_{m, W_m - 2} 
	= \frac{2}{W_m} \left[ \frac{(p_{\text{MLD}})^m}{1 - p_{\text{MLD}}}  b_{0, 0} + b_{m', 0} + (1 - Y)  b_{m'', 0} \right] , \notag \\
&\qquad\vdots \notag \\
1 &\cdot b_{m, 1} 
	= \frac{W_m - 1}{W_m} \left[ \frac{(p_{\text{MLD}})^m}{1 - p_{\text{MLD}}}  b_{0, 0} + b_{m', 0} + (1 - Y)  b_{m'', 0} \right] , \notag
\end{align}
from which we conclude that
\end{comment}
\begin{align}
b_{m, k} = \frac{W_m - k}{W_m} \bigg[ \frac{(p_{\text{MLD}})^m}{1 - p_{\text{MLD}}}  b_{0, 0} + b_{m', 0} &+ (1 - Y)  b_{m'', 0} \bigg] , \notag \\
	& 1 \leq k \leq W_m - 1 . 
	%\quad \textcolor{red}{... (9)} 
	\label{eq:IV(9)}
\end{align}
Then, the state transitions to/from the state $(m',0)$ gives
\begin{align}
b_{m', 0} 
	&= \frac{p_{\text{MLD}}}{W_m} \cdot X_{\text{MLD}} \cdot b_{m - 1, 0} 
		+ \frac{p_{\text{MLD}}}{W_m} \cdot X_{\text{MLD}} \cdot b_{m, 0} \notag\\
	&+ \frac{1}{W_m} \cdot b_{m'', 0} %\notag \\
	%&\quad
	+ \frac{1}{W_m} \cdot (1 - Y) \cdot b_{m'', 0} + X_{\text{MLD}} \cdot b_{m, 1} , \notag
\end{align}
to which we can apply Eq. \eqref{eq:IV(9)} to obtain
%\begin{align}
%\frac{W_m - 1}{W_m} \cdot b_{m', 0} &= \frac{X_{MLD}}{W_m} \cdot \frac{p^m}{1 - p} \cdot b_{0, 0} + \frac{1}{W_m} (1 - Y) \cdot b_{m'', 0} \notag \\
%	&\quad+ \frac{W_m - 1}{W_m} \cdot X_{MLD}  
%		\bigg[ \frac{p^m}{1 - p} \cdot b_{0, 0} + b_{m', 0} \notag \\
%	&\qquad\qquad\qquad\qquad+ (1 - Y) \cdot b_{m'', 0} \bigg] , \notag\\
%(1 - X_{MLD}) & \frac{W_{m-1}}{W_m} \cdot b_{m',0} 
%	= X_{MLD} \cdot \frac{p^m}{1 - p} \cdot b_{0,0} 
%		+\frac{(1 - Y)}{W_m} \cdot b_{m'',0} \notag \\
%	&\qquad\qquad\qquad\quad+ \frac{W_{m-1}}{W_m} X_{MLD} (1 - Y) \cdot b_{m'',0} . \notag \\
%\end{align}
\begin{align}
b_{m',0} &= \frac{X_{\text{MLD}}}{1 - X_{\text{MLD}}} \cdot \frac{W_m}{W_{m-1}} \cdot \frac{(p_{\text{MLD}})^m}{1 - p_{\text{MLD}}} b_{0,0} \notag\\
&+ \frac{1 - Y}{1 - X_{\text{MLD}}} \bigg( \frac{1}{W_{m-1}} \cdot b_{m'',0} 
	+ X_{\text{MLD}} \cdot b_{m',0} \bigg) . 
\label{eq:IV_b_mprime_0} 
%b_{m',0} &= \frac{X_{MLD}}{1 - X_{MLD}} \cdot \frac{W_m}{W_{m-1}} \cdot \frac{p^m}{1 - p} b_{0,0} + \frac{1}{W_{m-1}} \cdot \frac{1 - Y}{1 - X_{MLD}} b_{m'',0} \notag\\
%	&\quad + \frac{X_{MLD}}{1 - X_{MLD}} \cdot (1 - Y) \cdot b_{m',0} . 
%\label{eq:IV_b_mprime_0}
\end{align}
Also, by considering the state transitions to/from the state $(m'',0)$,
\begin{align}
b_{m'',0} &= (1 - X_{\text{MLD}}) \cdot (1 - \tau_{\text{MLD}}) \notag \\  &\quad \cdot \Big\{ \frac{p_{\text{MLD}}}{W_m} \cdot  \left( b_{m-1,0} + b_{m,0} \right) 
	+ b_{m,1} \Big\} , 
%b_{m'',0} = (1 - X_{\text{MLD}}) (1 - \tau_{\text{MLD}}) \Big\{
%\frac{p_{\text{MLD}}}{W_m} \left( b_{m-1,0} + b_{m,0} \right) 
%+ b_{m,1} \Big\} , \notag \\	
\end{align}
to which we can also apply Eq. \eqref{eq:IV(9)}, leading to
%\begin{align}
%b_{m'',0} =&\; \frac{p}{W_m} (1 - X_{MLD}) (1 - \tau) b_{m-1,0} \notag \\
%	&+ \frac{p}{W_m} (1 - X_{MLD}) (1 - \tau) b_{m,0} 
%	+ (1 - X_{MLD})(1 - \tau) b_{m,1} \notag \\
%	=&\; \frac{1}{W_m} (1 - X_{MLD}) (1 - \tau) \cdot \frac{p^m}{1 - p} \cdot b_{0,0} \notag \\
%	&+ (1 - X_{MLD}) (1 - \tau) \cdot \frac{W_{m-1}}{W_m} \notag \\ 
%	&\quad\cdot \left[ \frac{p^m}{1 - p} \cdot b_{0,0}+ b_{m',0} + (1 - Y) \cdot b_{m'',0} \right] , \notag\\
%\bigg( 1 &- \frac{W_{m-1}}{W_m} (1 - X_{MLD}) (1 - \tau) (1 - Y) \bigg) \cdot b_{m'' , 0} \notag\\
%	&= (1 - X_{MLD}) (1 - \tau) \frac{p^m}{1 - p} \cdot b_{0,0} 
%		+ X_{MLD} (1 - \tau) \frac{p^m}{1 - p} \cdot b_{0,0} \notag\\
%	&\quad+ \frac{1}{W_m} (1 - \tau) (1 - Y) \cdot b_{m'',0} \notag\\
%	&\quad+ \frac{W_{m-1}}{W_m} X_{MLD} (1 - \tau) (1 - Y) \cdot b_{m'',0} . \notag
%\end{align}
%\textcolor{red}{[HK: in the above, what is the relationship among the six equations?]}
\begin{align}
b_{m'',0} = \frac{ 1 - \tau_{\text{MLD}} }{1 - (1 - \tau_{\text{MLD}})(1 - Y)} \cdot \frac{(p_{\text{MLD}})^m}{1 - p_{\text{MLD}}} \cdot b_{0,0} . 
	%\quad \textcolor{red}{... (11)} 
	\label{eq:IV(11)} 
\end{align}
%\textcolor{red}{[HK: how do we obtain the equation below? -- (61)+(63)]}
Then, applying Eq. \eqref{eq:IV(11)} to Eq. \eqref{eq:IV_b_mprime_0} provides %the above provides
\begin{align}
b_{m',0} 
&= \bigg[ \frac{X_{\text{MLD}}}{1 - X_{\text{MLD}}} \cdot \frac{W_m}{W_{m-1}} 
	+ \frac{1}{W_{m-1}} \cdot \textcolor{black}{\lambda_3} \bigg] 
		 \frac{(p_{\text{MLD}})^m}{1 - p_{\text{MLD}}} \cdot b_{0,0} \notag\\
	&\quad+ X_{\text{MLD}} \cdot \textcolor{black}{\lambda_3} \cdot \frac{(p_{\text{MLD}})^m}{1 - p_{\text{MLD}}} \cdot b_{0,0} . %\notag\\
	%&\quad \textcolor{red}{... (12)} 
	\label{eq:IV(12)}
%b_{m',0} 
%&= \bigg[ \frac{X_{MLD}}{1 - X_{MLD}} \cdot \frac{W_m}{W_{m-1}} 
%	+ \frac{1}{W_{m-1}} \cdot \frac{1 - Y}{1 - X_{MLD}} \notag\\
%	&\quad\cdot \frac{1 - \tau}{1 - (1 - \tau)(1 - Y)} \bigg] 
%		\cdot \frac{p^m}{1 - p} \cdot b_{0,0} \notag\\
%	&\quad+ \frac{X_{MLD}}{1 - X_{MLD}} (1 - Y) \frac{1 - \tau}{1 - (1 - \tau)(1 - Y)} \frac{p^m}{1 - p} b_{0,0} . \notag \\
%	&\quad \textcolor{red}{... (12)} \label{eq:IV(12)}
%b_{m',0} 
%&= \bigg[ \frac{X_{MLD}}{1 - X_{MLD}} \cdot \frac{W_m}{W_{m-1}} 
%	\cdot \frac{p^m}{1 - p} + \frac{1}{W_{m-1}} \cdot \frac{1 - Y}{1 - X_{MLD}} \notag\\
%	&\qquad\cdot \frac{1 - \tau}{1 - (1 - \tau)(1 - Y)} \cdot \frac{p^m}{1 - p} \bigg] b_{0,0} \notag\\
%	&\qquad+ \frac{X_{MLD}}{1 - X_{MLD}} (1 - Y) \frac{1 - \tau}{1 - (1 - \tau)(1 - Y)} \cdot \frac{p^m}{1 - p} \cdot b_{0,0} \notag\\
\end{align}
By combining Eqs. \eqref{eq:IV(9)}, \eqref{eq:IV(11)}, and \eqref{eq:IV(12)}, %(9), (11), and (12),
\begin{align}
b_{m,k} &= \frac{W_m - k}{W_m} \cdot \frac{(p_{\text{MLD}})^m}{1 - p_{\text{MLD}}} \cdot \bigg[ 1 + \frac{X_{\text{MLD}}}{1 - X_{\text{MLD}}} \cdot \frac{W_m}{W_{m-1}} \notag\\
	&\quad+ \frac{W_m}{W_{m-1}} \cdot \textcolor{black}{\lambda_3} \bigg] \cdot b_{0,0} , \quad 1 \leq k \leq W_m - 1 , \notag\\
%b_{m,k} &= \frac{W_m - k}{W_m} \cdot \frac{p^m}{1 - p} \cdot \bigg[ 1 + \frac{X_{MLD}}{1 - X_{MLD}} \cdot \frac{W_m}{W_{m-1}} \notag\\
%	&\quad+ \frac{W_m}{W_{m-1}} \cdot \frac{1 - Y}{1 - X_{MLD}} \cdot \frac{1 - \tau}{1 - (1 - \tau)(1 - Y)} \bigg] \cdot b_{0,0} , \notag\\ 
%		&\qquad\qquad\qquad\qquad\qquad\qquad\qquad\quad 1 \leq k \leq W_m - 1 , \notag\\
b_{m,0} &= \frac{(p_{\text{MLD}})^m}{1 - p_{\text{MLD}}} \cdot b_{0,0} , \quad k = 0 . \notag
\end{align}

%\textcolor{red}{For $1 \leq i \leq m - 1$, }
Following the procedures above similarly, the state transitions at each state for $1 \leq i \leq m-1$ provide
%Next, we try to derive $b_{i,k}$'s for $1 \leq i \leq m-1$, where
%Each state for $1 \leq i \leq m-1$ provides}
\begin{comment}
\textcolor{magenta}{	%%%%%%%%%%%%%%%%%%%%%%%%
\begin{align}
1 \cdot b_{i,W_i-1} &= \frac{p_{\text{MLD}}}{W_i} \cdot b_{i-1,0} + \frac{1}{W_i} \cdot b_{i,0} + \frac{1}{W_i} (1 - Y) \cdot b_{i'',0} \notag \\[10pt]
	&= \frac{1}{W_i} \left[ (p_{\text{MLD}})^i \cdot b_{0,0} + b_{i,0} + (1 - Y) \cdot b_{i'',0} \right] , \notag\\%[10pt]
	&\vdots \notag \\[10pt]
1 \cdot b_{i,1} &= \frac{W_i - 1}{W_i} \left[ (p_{\text{MLD}})^i \cdot b_{0,0} + b_{i,0} + (1 - Y) \cdot b_{i'',0} \right] . \notag %\\[10pt]
\end{align}
from which we obtain} %%%%%%%%%%%%%%%%%%%%%%%%%%%%%%%%%%%%%%%%%%%%
\end{comment}
\begin{align}
b_{i,k} &= \frac{W_i - k}{W_i} \left\{ (p_{\text{MLD}})^i \cdot b_{0,0} + b_{i',0} + (1 - Y) \cdot b_{i'',0} \right\} , \notag\\ 
&\qquad\qquad\qquad\qquad\qquad\qquad\quad 
	1 \leq k \leq W_i - 1 . 
	%\textcolor{red}{... (13)} 
	\label{eq:IV(13)}
\end{align}

%\textcolor{red}{[HK: in the below, how do we obtain \& what is the relationship among the four equations?]}
Next, the state transitions to/from the state $(i',0)$ provide
\begin{align}
b_{i',0} &= \frac{p_{\text{MLD}}}{W_i} \cdot X_{\text{MLD}} \cdot b_{i-1,0} + \frac{1}{W_i} \cdot b_{i',0} \notag\\
	&\quad+ \frac{1}{W_i} \cdot (1 - Y) \cdot b_{i'',0} + X_{\text{MLD}} \cdot b_{i,1} , \notag
%\frac{W_i - 1}{W_i} \cdot b_{i',0} &= \frac{1}{W_i} \cdot X_{MLD} \cdot p^i \cdot b_{0,0} + \frac{1}{W_i} \cdot (1 - Y) \cdot b_{i'',0} \notag\\%[6pt]
%	&\quad+ \frac{W_i - 1}{W_i} \cdot X_{MLD} \cdot p^i \cdot b_{0,0} 
%		+ \frac{W_i - 1}{W_i} \cdot X_{MLD} \cdot b_{i',0} \notag\\
%	&\quad+ \frac{W_i - 1}{W_i} \cdot X_{MLD} \cdot (1 - Y) \cdot b_{i'',0} \notag\\%[8pt]
%\frac{W_i - 1}{W_i} \cdot (1-X_{MLD}) \cdot b_{i',0} &= X_{MLD} \cdot p^i \cdot b_{0,0} 
%	+ \frac{1}{W_i} \cdot (1 - Y) \cdot b_{i'',0} \notag\\%[6pt]
%	&\quad + \frac{W_i - 1}{W_i} \cdot X_{MLD} \cdot (1 - Y) \cdot b_{i'',0} \notag\\%[12pt]
\end{align}
to which we can apply Eq. \eqref{eq:IV(13)} to obtain
\begin{align}
b_{i',0} &= 
	\frac{1}{W_{i-1}} \cdot \frac{1 - Y}{1 - X_{\text{MLD}}} \cdot b_{i'',0} 
	+ \frac{X_{\text{MLD}} \cdot (1 - Y)}{1 - X_{\text{MLD}}} \cdot b_{i'',0} \notag\\
	&\quad+ \frac{X_{\text{MLD}}}{1 - X_{\text{MLD}}} \cdot \frac{W_i}{W_{i-1}} \cdot (p_{\text{MLD}})^i \cdot b_{0,0} .
	%\quad \textcolor{red}{... (16)} 
	\label{eq:IV(16)}
%b_{i',0} &= \frac{X_{MLD}}{1 - X_{MLD}} \cdot \frac{W_i}{W_{i-1}} \cdot p^i \cdot b_{0,0} 
%	+ \frac{1}{W_{i-1}} \cdot \frac{1 - Y}{1 - X_{MLD}} \cdot b_{i'',0} \notag\\
%	&\quad + \frac{X_{MLD}}{1 - X_{MLD}} \cdot (1 - Y) \cdot b_{i'',0}. 
%	\quad \textcolor{red}{... (16)} 
%	\label{eq:IV(16)}
\end{align}
In addition, the state transitions to/from the state $(i'',0)$ provide
\begin{comment}
\begin{align}
b_{i'',0} = (1 - X_{\text{MLD}})(1 - \tau_{\text{MLD}}) \bigg( \frac{1}{W_i} \cdot p_{\text{MLD}} \cdot b_{i-1,0} + b_{i,1} \bigg) .
	%\quad \textcolor{red}{... (14)} 
	\label{eq:IV(14)}
\end{align}
Then, applying Eq. \eqref{eq:IV(13)} to Eq. \eqref{eq:IV(14)} leads to
\end{comment}
%\begin{align}
%b_{i'',0}	&= (1 - X_{MLD})(1 - \tau) \frac{p}{W_i} \cdot b_{0,0} 
%	+ (1 - X_{MLD})(1 - \tau) \cdot \frac{W_i - 1}{W_i} \notag\\
%	&\quad\cdot \textcolor{green}{\left[ p^i \cdot b_{0,0} + b_{i,0} + (1-Y) \cdot b_{i'',0} \right]} \notag \\
%\left[ \right. 1 &- \frac{W_i - 1}{W_i} \cdot (1 - X_{MLD})(1 - \tau)(1 - Y) \left. \right] \cdot b_{i'',0} \notag\\%[8pt]
%	&= (1 - X_{MLD})(1 - \tau) \cdot p^i \cdot b_{0,0} + X_{MLD} \cdot (1 - \tau) \cdot p^i \cdot b_{0,0} \notag\\%[6pt]
%	&\quad + \frac{1}{W_i} \cdot (1 - \tau)(1 - Y) \cdot b_{i'',0} \notag\\%[6pt]
%	&\quad + \frac{W_i - 1}{W_i} \cdot X_{MLD} \cdot (1 - \tau)(1 - Y) \cdot b_{i'',0} , \notag %\\[8pt]
%\end{align}
%\textcolor{red}{[HK: in the above, what is the relationship among the two equations?]}
%and thus we have
\begin{align}
b_{i'',0} &= \frac{1 - \tau_{\text{MLD}}}{1 - (1 - \tau_{\text{MLD}})(1 - Y)} \cdot (p_{\text{MLD}})^i \cdot b_{0,0} . 
	%\quad \textcolor{red}{... (15)} 
	\label{eq:IV(15)}
\end{align}
Then by applying Eq. \eqref{eq:IV(15)} to Eq. \eqref{eq:IV(16)}, we obtain
\begin{align}
b_{i',0} &= \left[ 
	\frac{X_{\text{MLD}}}{1 - X_{\text{MLD}}} \frac{W_i}{W_{i-1}} 
	+ \frac{\textcolor{black}{\lambda_3}}{W_{i-1}}   
	+ X_{\text{MLD}} \textcolor{black}{\lambda_3} \right] (p_{\text{MLD}})^i  b_{0,0}. 
	%\quad \textcolor{red}{... (17)} 
	\label{eq:IV(17)} 
%b_{i',0} &= \left[ 
%	\frac{X_{MLD}}{1 - X_{MLD}} \cdot \frac{W_i}{W_{i-1}} 
%	+ \frac{1}{W_{i-1}} \cdot \frac{1 - Y}{1 - X_{MLD}} \cdot \frac{1 - \tau}{1 - (1 - \tau)(1 - Y)}  \right. \notag\\%[6pt]
%&\left. \quad + \frac{X_{MLD}}{1 - X_{MLD}} \cdot (1 - Y) 
%	\cdot \frac{1 - \tau}{1 - (1 - \tau)(1 - Y)} \right] \cdot p^i \cdot b_{0,0}. \quad \textcolor{red}{... (17)} \label{eq:IV(17)}
\end{align}
By combining Eqs. \eqref{eq:IV(13)}, \eqref{eq:IV(15)}, and \eqref{eq:IV(17)},
\begin{align}
b_{i,k} &= \frac{W_i - k}{W_i} \bigg\{ 
	1 	+ \frac{W_i}{W_{i-1}} \bigg( 
		\frac{X_{\text{MLD}}}{1 - X_{\text{MLD}}}  
		+ \textcolor{black}{\lambda_3} \bigg)
	\bigg\}  (p_{\text{MLD}})^i b_{0,0} , \notag\\
	&\qquad\qquad\qquad\qquad\qquad\qquad\qquad\qquad 1 \leq k \leq W_i - 1 , \notag\\
%b_{i,k} &= \frac{W_i - k}{W_i} \left[ 1 + \frac{X_{MLD}}{1 - X_{MLD}} \cdot \frac{W_i}{W_{i-1}} \right. \notag\\%[6pt]
%	&\quad \left. + \frac{1 - Y}{1 - X_{MLD}} \cdot \frac{W_i}{W_{i-1}} \cdot \frac{1 - \tau}{1 - (1 - \tau)(1 - Y)} \right] \cdot p \cdot b_{0,0} , \notag\\%[8pt]
%	&\qquad\qquad\qquad\qquad\qquad\qquad\qquad 1 \leq k \leq W_i - 1 , \notag\\%[10pt]
b_{i,0} &= (p_{\text{MLD}})^i \cdot b_{0,0} , \quad k = 0 . 
%\quad \textcolor{red}{... (18)} 
\label{eq:IV(18)}
\end{align}
Since the sum of stationary probabilities should be 1, we get
\begin{align}
1 &= 
	\sum_{i=0}^{m} \sum_{k=0}^{W_i - 1} b_{i,k} 
	+ \sum_{i'=0'}^{m'} b_{i',0} 
	+ \sum_{i''=0''}^{m''} b_{i'',0} 
	= \sum_{k=0}^{W_0 - 1} b_{0,k} \notag\\
  	&\quad+ \sum_{k=0}^{W_m - 1} b_{m,k} 
  	+ \sum_{i=1}^{m-1} \sum_{k=0}^{W_i - 1} b_{i,k} 
  	+ \sum_{i'=0'}^{m'} b_{i',0} 
  + \sum_{i''=0''}^{m''} b_{i'',0}. 
  %\quad \textcolor{red}{... (19)} 
  \label{eq:IV(19)}
\end{align}
By applying Eqs. \eqref{eq:IV(6)} through \eqref{eq:IV(18)} to the above, we obtain %and reducing it as much as possible, we obtain
\begin{align}
1 &= \bigg[ \frac{1}{1-p_{\text{MLD}}} + \bigg\{ \frac{W_0-1}{2} + \frac{X_{\text{MLD}}}{1-X_{\text{MLD}}} \cdot \frac{W_0}{2} + \frac{W_0}{2} \cdot \textcolor{black}{\lambda_3} \bigg\} \notag\\
&\;\;+ \bigg\{ \frac{W_m-1}{2} + \frac{X_{\text{MLD}}}{1-X_{\text{MLD}}} \cdot \frac{W_m}{2} + \frac{W_m}{2} \cdot \textcolor{black}{\lambda_3} \bigg\}  \frac{(p_{\text{MLD}})^m}{1-p_{\text{MLD}}} \notag\\
&\;\;+ \sum_{i=1}^{m-1} \sum_{k=1}^{W_i-1} \bigg\{ 
	\frac{W_i-k}{W_i} + \frac{W_i-k}{W_i-1} \bigg( 
		\frac{X_{\text{MLD}}}{1-X_{\text{MLD}}} + \textcolor{black}{\lambda_3} \bigg)
	\bigg\} p^i \notag\\
&\;\;+ \sum_{i=0}^{m-1} \bigg\{ 
	\frac{1}{W_i-1} \bigg( \frac{X_{\text{MLD}} \cdot W_i}{1-X_{\text{MLD}}} 
	+ \textcolor{black}{\lambda_3} \bigg)
	+ X_{\text{MLD}} \textcolor{black}{\lambda_3} 
	\bigg\} (p_{\text{MLD}})^i \notag\\
&\;\;+ \bigg\{ 
	\frac{1}{W_m-1} \bigg( \frac{X_{\text{MLD}} \cdot W_m}{1-X_{\text{MLD}}} 
	+ \textcolor{black}{\lambda_3} \bigg)
	+ X_{\text{MLD}} \textcolor{black}{\lambda_3} 
	\bigg\} \frac{(p_{\text{MLD}})^m}{1-p_{\text{MLD}}} \notag\\
&\;\;+ \frac{\textcolor{black}{\lambda_3}}{1-p_{\text{MLD}}} \cdot \frac{1-X_{\text{MLD}}}{1-Y}
 \bigg] \cdot b_{0,0} . %\notag\\
	%\quad \textcolor{red}{... (20)} 
	\label{eq:IV(20)}
\end{align}
\subsection{Derivation of transmit probabilities}

We are now ready to derive the transmit probability of a non-AP MLD STA towards the AP MLD, denoted by $\tau_\text{MLD}$, which is given as
\begin{align}
\tau_{\text{MLD}} = \sum_{i=0}^{m} b_{i,0} = \frac{b_{0,0}}{1 - p_{\text{MLD}}},
\end{align}
where $b_{0,0}$ can be obtained 
from Eq. \eqref{eq:IV(20)}.

%% file: analysis.tex
\subsection{Transmit probabilities of SLD} %legacy Wi-Fi devices}
\label{subsec:tau_leagcy}

For legacy devices, the transmit probability $\tau_{\text{SLD}}$ can be derived from the Bianchi model in \cite{bianchi2000performance,magrin2023performance}.
Considering that $\tau_{\text{SLD}}$ is influenced by the destination of an AP MLD's transmitted packet (Case 1 or Case 2), we derive $\tau_{\text{SLD}}$ for each case as follows:
\begin{align}
\tau_{\text{SLD}}^{I} &= \left[ 
	\frac{1 - p_{\text{SLD}}^{I} - p_{\text{SLD}}^{I} (2p_{\text{SLD}}^{I})^m}{1-2p_{\text{SLD}}^{I}} \cdot \frac{(CW_{min}^{\text{SLD}}+1)}{2} + \frac{1}{2}
	\right]^{-1} , \notag\\
%\end{align}
%\begin{align}
\tau_{\text{SLD}}^{II} &= \left[ 
	\frac{1 - p_{\text{SLD}}^{II} - p_{\text{SLD}}^{II} (2p_{\text{SLD}}^{II})^m }{1-2p_{\text{SLD}}^{II}} \cdot \frac{(CW_{min}^{\text{SLD}}+1)}{2} + \frac{1}{2}
	\right]^{-1} , \notag
\end{align}
where $CW_{min}^{\text{SLD}}$ is the minimum contention window size employed by the legacy devices.

%-------------------------------------------------------
\subsection{Collision probabilities of AP MLD, non-AP MLD, SLD}
\label{subsec:collision_prob}

We derive the {\it collision} probability $p$, per device and per slot, using the derived $\tau$'s for Case 1 and Case 2. %that a device would experience a collision in a randomly chosen slot.
%\textcolor{green}{[SH: Considering the structure of the paper, it would be better to introduce the notation for $p$ at the end of Section 3.]}
%For example, $p_{SLD}^I$ is computed as 
%\textcolor{blue}{Such collisions occur when one type of device attempts to transmit while one or more coexisting devices also attempt to transmit.}
%
For Case 1, the collision probabilities of an AP MLD, a non-AP MLD, and an SLD are denoted by $p_{\text{AP}}^{\text{SLD}}$, $p_{\text{MLD}}^{I}$, and $p_{\text{SLD}}^I$, respectively, which can be derived as
\begin{align}
p_{\text{AP}}^{\text{SLD}} &= 
	1 - ( 1 - \tau_{\text{SLD}}^I 	)^{N_{\text{SLD}}} ( 1 - \tau_{\text{MLD}}^I 	)^{N_{\text{MLD}}} , \notag\\
p_{\text{MLD}}^{I} &= 
	1 - ( 1 - \tau_{\text{AP}}^{\text{SLD}} ) ( 1 - \tau_{\text{SLD}}^I 	)^{N_{\text{SLD}}} ( 1 - \tau_{\text{MLD}}^I 	)^{N_{\text{MLD}} - 1}, \notag\\
p_{\text{SLD}}^I &= 
	1 - ( 1 - \tau_{\text{AP}}^{\text{SLD}} ) ( 1 - \tau_{\text{SLD}}^I )^{N_{\text{SLD}} - 1} ( 1 - \tau_{\text{MLD}}^I 	)^{N_{\text{MLD}}} . \notag
\end{align}
Similarly, the collision probabilities of an AP MLD, a non-AP MLD, and an SLD in Case 2 are denoted by $p_{\text{AP}}^{\text{MLD}}$, $p_{\text{MLD}}^{II}$, and $p_{\text{SLD}}^{II}$, respectively, which are derived as
\begin{align}
p_{\text{AP}}^{\text{MLD}} &= 
	1 - ( 1 - \tau_{\text{SLD}}^{II} )^{N_{\text{SLD}}} ( 1 - \tau_{\text{MLD}}^{II} )^{N_{\text{MLD}}}, \notag\\
p_{\text{MLD}}^{II} &= 1 - ( 1 - \tau_{\text{AP}}^{\text{MLD}} ) ( 1 - \tau_{\text{SLD}}^{II} )^{N_{\text{SLD}}} ( 1 - \tau_{\text{MLD}}^{II} 	)^{N_{\text{MLD}} - 1}, \notag\\
p_{\text{SLD}}^{II} &= 1 - ( 1 - \tau_{\text{AP}}^{\text{MLD}} ) ( 1 - \tau_{\text{SLD}}^{II} )^{N_{\text{SLD}} - 1} ( 1 - \tau_{\text{MLD}}^{II} 	)^{N_{\text{MLD}}}. \notag
\end{align}
%\begin{align}
%\label{eq:p_legacy_1}
%p_{\scriptscriptstyle SLD}^I = 1 - (1-\tau_{\scriptscriptstyle AP}^{\scriptscriptstyle SLD}) (1-\tau_{\scriptscriptstyle SLD}^I)^{N_{\scriptscriptstyle SLD}-1} (1-\tau_{\scriptscriptstyle MLD}^I)^{N_{\scriptscriptstyle MLD}} . \notag
%\end{align}
%Derivation of other collision probabilities is also omitted due to space limit, but they can be obtained similarly.

Next, we derive the probability of every possible event occurring in a random slot, using $\tau$'s and $p$'s.
In Case 1, the probability $p_{idle}^{I}$ that no one transmits in a slot is derived as 
\begin{align}
%\label{eq:b_1}
p_{idle}^{I} = (1-\tau_{\text{AP}}^{\text{SLD}}) (1-\tau_{\text{SLD}}^I)^{N_{\text{SLD}}} (1-\tau_{\text{MLD}}^I)^{N_{\text{MLD}}} . \notag
\end{align}
The probability $\tau_{1}^{I}$ that the AP MLD alone transmits a packet destined to a legacy STA is derived as 
\begin{align}
%\label{eq:a_1}
\tau_{1}^{I} = \tau_{\text{AP}}^{\text{MLD}} \cdot (1-p_{\text{AP}}^{\text{SLD}}) . \notag
\end{align}
The probability $\tau_{2}^{I}$ that a legacy device alone transmits (to the AP) is derived as 
\begin{align}
%\label{eq:a_2}
\tau_{2}^{I} = N_{\text{SLD}} \cdot \tau_{\text{SLD}}^I \cdot (1-p_{\text{SLD}}^I) . \notag
\end{align}
The probability $\tau_{3}^{I}$ that a non-AP MLD alone transmits (to the AP) is derived as 
\begin{align}
%\label{eq:a_3}
\tau_{3}^{I} = N_{\text{MLD}} \cdot \tau_{\text{MLD}}^I \cdot (1-p_{\text{MLD}}^I) . \notag
\end{align}
The collision probability $p_{c,1}^{I}$ {\it between} AP MLD's DL transmission to a legacy device {\it and} one or more UL transmissions from non-AP MLDs and/or legacy devices is derived as
\begin{align}
%\label{eq:c_1}
p_{c,1}^{I} = \tau_{\text{AP}}^{\text{SLD}} \cdot p_{\text{AP}}^{\text{SLD}} . \notag
\end{align}
The collision probability $p_{c,2}^{I}$ between non-AP MLDs and legacy devices is derived as
\begin{align}
%\label{eq:c_2}
p_{c,2}^{I} = 1 - p_{idle}^{I} - (\tau_{1}^{I}+\tau_{2}^{I}+\tau_{3}^{I}) - p_{c,1}^{I} . \notag
\end{align}
In Case 2, 
the aforementioned $p_{idle}^{I}$, $\tau_{2}^{I}$, $\tau_{3}^{I}$, $p_{c,1}^{I}$, and $p_{c,2}^{I}$ are 
re-denoted by $p_{idle}^{II}$, $\tau_{2}^{II}$, $\tau_{3}^{II}$, $p_{c,1}^{II}$, and $p_{c,2}^{II}$, respectively, where they can be derived similar to Case 1 but using Case 2's $\tau$ and $p$.
%The probability $p_{idle}^{II}$ of no one transmitting, the probability $\tau_{2}^{II}$ of only one legacy device transmitting, the probability $\tau_{3}^{II}$ of only one non-AP MLD transmitting, the probability $p_{c,1}^{II}$ of the AP MLD experiencing collision, and the probability $p_{c,2}^{II}$ of collision between non-AP MLDs and legacy devices can be derived similarly to Case 1, while using Case 2's $\tau$ and $p$.
On the other hand, when the AP MLD alone transmits a packet to a non-AP MLD, we need to consider two different cases.  
%The distinction, however, arises when the AP MLD successfully transmits a packet to a non-AP MLD.
%When the AP MLD alone transmits a packet to non-AP MLDs, 
First, the AP MLD can transmit either through a single link to the non-AP MLD or through both links to a couple of non-AP MLDs, with probability $\tau_{1a}^{II}$, which is derived as
\begin{align}
%\label{eq:a_4}
\tau_{1a}^{II} = \left( 1-\frac{1}{N_{\text{MLD}}} \cdot \tau_{\text{AP}}^{\text{MLD}} \right) \cdot \tau_{\text{AP}}^{\text{MLD}} , \notag
\end{align}
where ${1}/{N_{\text{MLD}}}$ represents the probability that in case both links are used, the destination of the two packets is the same non-AP MLD.
Second, the AP MLD may transmit alone through both links while its destination is the same non-AP MLD, with probability $\tau_{1b}^{II}$, which is derived as 
\begin{align}
%\label{eq:a_5}
\tau_{1b}^{II} = \frac{1}{N_{\text{MLD}}} \cdot \tau_{\text{AP}}^{\text{MLD}} \cdot \tau_{\text{AP}}^{\text{MLD}} . \notag
\end{align}
Then, the other probabilities can be expressed as
\begin{align}
p_{idle}^{II} &= \left(1-\tau_{\text{AP}}^{\text{MLD}}\right) \cdot\left(1-\tau_{\text{SLD}}^{I I}\right)^{N_{\text{SLD}}} \cdot\left(1-\tau_{\text{MLD}}^{II}\right)^{N_{\text{MLD}}} , \notag\\
\tau_{2}^{II} &= N_{\text{SLD}} \cdot \tau_{\text{SLD}}^{II} \cdot\left(1-\tau_{A P}^{M L D}\right) \cdot\left(1-\tau_{\text{SLD}}^{II}\right)^{N_{\text{SLD}}-1} \notag \\
	&\quad\cdot\left(1-\tau_{M D}^{II}\right)^{N_{M L D}} , \notag\\
\tau_{3}^{II} &= N_{\text{MLD}} \cdot \tau_{\text{MLD}}^{II} \cdot\left(1-\tau_{\text{AP}}^{\text{MLD}}\right) \cdot\left(1-\tau_{\text{SLD}}^{II}\right)^{N_{\text{SLD}}} \notag\\
	&\quad\cdot\left(1-\tau_{\text{MLD}}^{II}\right)^{N_{\text{MLD}}-1} , \notag\\
p_{c,1}^{II} &= \tau_{\text{AP}}^{\text{MLD}} \cdot p_{\text{AP}}^{\text{MLD}} , \notag\\
p_{c,2}^{II} &= 1-\left(p_{idle}^{II}+b_3\right)-\left(\tau_{1a}^{II}+\tau_{1b}^{II}+\tau_{2}^{II}+\tau_{3}^{II}\right)-p_{c,1}^{II} . \notag
\end{align}

%-------------------------------------------------------
\subsection{Duration of each event in a random slot}

Each aforementioned event has a corresponding time duration.
We briefly explain how these durations are derived.
We assume that all transmissions are based on Aggregated MPDUs (A-MPDU), and all devices use the same modulation coding scheme (MCS) index.
Then, the time $T_{data}$ required for data transmission is given as 
\begin{align}
\label{T_data}
T_{data} = T_{PHY} + \left\lceil {L_{AMPDU}}/{r_{SU}} \right\rceil \cdot \sigma
\end{align}
where $T_{PHY}$ is the duration of PHY preamble and header, $L_{AMPDU}$ is the length of both DL and UL packets, $r_{SU}$ is the data rate (bits/symbol), and $\sigma$ is the OFDM symbol duration given by \cite{magrin2023performance}.
With $T_{data}$ given, the time $T_{success}$ to take for a successful transmission is derived as 
\begin{align}
T_{success} = T_{DATA} + SIFS + T_{ACK} + SIFS , 
\end{align}
where $T_{ACK}$ is the ACK frame duration. 
Then, the duration $T_{collision}$ of collision is derived as 
\begin{align}
T_{collision} = T_{DATA} + SIFS. 
\end{align}
As a result, the time duration for each event is as follows: $T_{\tau_{1}^{I}} = T_{\tau_{2}^{I}} = \cdots = T_{\tau_{3}^{II}} = T_{success}$ and $T_{p_{c,1}^{I}} = T_{p_{c,2}^{I}} = \cdots = T_{p_{c,2}^{II}} = T_{collision}$. 
Then, we can calculate the average duration $\phi$ considering all the events including successful transmission, collision, and an empty slot as follows:
\begin{align}
\label{eq:phi_case_1}
\phi_{case_1} &= p_{idle}^{I} \cdot T_{empty}  + (\tau_{1}^{I} + \tau_{2}^{I} + \tau_{3}^{I}) (T_{success} + T_{empty}) \nonumber \\
&\quad+ (p_{c,1}^{I}+p_{c,2}^{I}) (T_{collision} + T_{empty}) , \\
\label{eq:phi_case_2}
\phi_{case_2} &= p_{idle}^{II} \cdot T_{empty} + (\tau_{1a}^{II} + \tau_{1b}^{II} + \tau_{2}^{II} + \tau_{3}^{II}) (T_{success} \nonumber \\
&\quad+ T_{empty}) + (p_{c,1}^{II}+p_{c,2}^{II}) (T_{collision} + T_{empty}) ,
\end{align}
where $T_{empty}$ is the empty slot duration.

%-------------------------------------------------------
\subsection{Derivation of $X_{\text{AP}}$, $X_{\text{MLD}}$, and $Y$}
\label{subsec:derivation_x_y}

In Section \ref{sec:ap_mc} and Section \ref{sec:nonap_mc}, the definitions of $X_{\text{AP}}$, $X_{\text{MLD}}$ and $Y$ were provided when deriving $\tau_{\text{AP}}$ and $\tau_{\text{MLD}}$.
We hereby derive $X_{\text{AP}}$, $X_{\text{MLD}}$, and $Y$, by utilizing the parameters derived earlier in this section.

First, remind that $X_{\text{AP}}$ is the probability that from the viewpoint of a certain link belonging to the AP MLD, the other link is busy due to the transmission either from non-AP MLD or legacy STA.
%This is in fact the proportion of time within a slot duration, excluding the periods when no one transmits, when only the AP MLD transmits, and when collisions occur during the AP MLD's transmission.
Accordingly, $X_{\text{AP}}$ can be obtained by considering the proportion of time in a slot where the channel is busy due to the transmission from either non-AP MLD or legacy STA.
For Case 1, the duration $\phi_{\text{busy}}^{\text{AP},I}$ per slot when the other link is busy due to the transmission by non-AP MLD or legacy STA is given as
\begin{align}
\phi_{\text{busy}}^{\text{AP},I} &= 
	\phi_{\text{case}_1} 
	- T_{\text{idle}}^I 
	- \tau_{\text{AP}}^{\text{SLD}} \cdot (1 - p_{\text{AP}}^{\text{SLD}}) \cdot T_{\text{success}} \notag \\
	&\quad- \tau_{\text{AP}}^{\text{SLD}} \cdot p_{\text{AP}}^{\text{SLD}} \cdot T_{\text{collision}} , \\
%\end{align}
%\begin{align}
T_{\text{idle}}^{I} &= p_{idle}^{I} \cdot T_{empty} , 
\end{align}
where $T_{\text{idle}}^I$ implies the period that the channel is idle per slot in Case 1.
Similarly, $\phi_{\text{busy}}^{\text{AP},II}$ and $T_{\text{idle}}^{II}$ for Case 2 are given as
\begin{align}
\phi_{\text{busy}}^{\text{AP},II} &= 
	\phi_{\text{case}_2} 
	- T_{\text{idle}}^{II} 
	- \tau_{\text{AP}}^{\text{MLD}} \cdot (1 - p_{\text{AP}}^{\text{MLD}}) \cdot T_{\text{success}} \notag \\
	&\quad- \tau_{\text{AP}}^{\text{MLD}} \cdot p_{\text{AP}}^{\text{MLD}} \cdot T_{\text{collision}} , \\
%\end{align}
%where 
%\begin{align}
T_{\text{idle}}^{II} &= p_{idle}^{II} \cdot T_{empty}. 
\end{align}
For a link, the other link is in Case 1 with probability $(1-\gamma)$ and in Case 2 with probability $\gamma$, and thus we obtain $X_{\text{AP}}$ as
\begin{align}
	X_{\text{AP}} &= 
	(1 - \gamma) \cdot 
	\frac{ \phi_{\text{busy}}^{\text{AP},I} }
	{ \phi_{\text{case}_1} } 
	+ \gamma \cdot 
	\frac{ \phi_{\text{busy}}^{\text{AP},II} }
	{ \phi_{\text{case}_2} } . \notag
\end{align}

Next, $X_{\text{MLD}}$ is the probability that from the perspective of a link belonging to a non-AP MLD, the other link is busy due to the transmission either from the AP MLD, legacy devices, or other non-AP MLDs.
%Following this, $X_{\text{MLD}}$ 
This can be obtained by %deriving the proportion of time the channel is busy due to the AP MLD, legacy devices and other non-AP MLDs during a slot, 
%This is determined as the proportion of time within a slot duration, 
excluding the periods when no transmission occurs, when a single non-AP MLD transmits, when the AP and a  non-AP MLD transmit simultaneously resulting in a collision, and when the AP does not transmit but a collision occurs between a non-AP MLD and other devices.
Specifically, in Case 1, the duration per slot $\phi_{\text{busy}}^{\text{MLD},I}$ when the other link is busy due to the transmission by the devices {\it excluding} one non-AP MLD is given as
\begin{align}
	\phi_{\text{busy}}^{\text{MLD},I} &= 
	\phi_{\text{case}_1} 
	- T_{\text{idle}}^I 
	- \tau_{\text{MLD}}^I \cdot (1-p_{\text{MLD}}^I) \cdot T_{\text{success}} \notag \\
	&\quad- \tau_{\text{AP}}^{\text{SLD}} \cdot \tau_{\text{MLD}}^I \cdot T_{\text{collision}}  
	- (1-\tau_{\text{AP}}^{\text{SLD}}) \cdot \tau_{\text{MLD}}^I \notag \\ 
	&\quad \cdot (1-(1-\tau_{\text{MLD}}^I)^{N-1} \cdot (1-\tau_{\text{legacy}}^I)^N) \cdot T_{\text{collision}}.
\end{align}
Similarly, $\phi_{\text{busy}}^{MLD,II}$ for Case 2 is derived as
\begin{align}
	\phi_{\text{busy}}^{\text{MLD},II} &= 
	\phi_{\text{case}_2} 
	- T_{\text{idle}}^{II} 
	- \tau_{\text{MLD}}^{II} \cdot (1-p_{\text{MLD}}^{II}) \cdot T_{\text{success}} \notag \\
	&\quad- \tau_{\text{AP}}^{\text{MLD}} \cdot \tau_{\text{MLD}}^{II} \cdot T_{\text{collision}}  
	- (1-\tau_{\text{AP}}^{\text{MLD}}) \cdot \tau_{\text{MLD}}^{II} \notag \\ 
	&\quad \cdot (1-(1-\tau_{\text{MLD}}^{II})^{N-1} \cdot (1-\tau_{\text{legacy}}^{II})^N) \cdot T_{\text{collision}}.
\end{align}
%Considering that from the perspective of one link, the other link is in Case 1 with probability $(1-\gamma)$ and in Case 2 with probability $\gamma$, 
Then we obtain $X_{\text{MLD}}$ as
\begin{align}
	X_{\text{MLD}} &= 
	(1 - \gamma) \cdot 
	\frac{ \phi_{\text{busy}}^{\text{MLD},I} }
	{ \phi_{\text{case}_1} } 
	+ \gamma \cdot 
	\frac{ \phi_{\text{busy}}^{\text{MLD},II} }
	{ \phi_{\text{case}_2} } . \notag
\end{align}

Finally, $Y$ denotes the probability that, within a non-AP MLD, one link's backoff counter remains at zero until the other link's backoff counter also reaches zero.
%\textcolor{blue}{To formulate $Y$, we must consider the conditions defined by its definition.}
%Within a non-AP MLD, we have been assumed that there are two links in an MLD.
%To aid understanding, let us assume the existence of Link 1 and Link 2.
Without loss of generality, assuming Link 1 is currently in $(i'',0)$, suppose Link 2's backoff counter is $j$ ($j \neq 0$). %\textcolor{red}{$b_{i,j}$} (where $j \neq 0$), 
Then, if both Link 1 and Link 2 remain idle for $j$ slots, Link 2's backoff counter will reach zero thus allowing simultaneous transmissions at both links (which corresponds to $(i'',0) \rightarrow (i,0)$ transition in Fig. \ref{fig:non_AP_MLD}).
%Considering that Link 2 falls under Case 1 with probability $1-\gamma$ and under Case 2 with probability $\gamma$ from Link 1's perspective, $Y$ is then formulated accordingly.
Therefore, when Link 1 is in Case 1, $Y_{{case}_1}$ is derived as
\begin{align}
Y_{{case}_1} &= \sum_{i''=0}^{m} \sum_{j=1}^{W_0 -1} \sum_{i=0}^{m} b_{i'',0}^{I} \cdot (p_{idle}^{I})^j \notag\\ 
	&\quad\cdot \left\{ (1-\gamma) \cdot b_{i,j}^{I} \cdot (p_{idle}^{I})^j + \gamma \cdot b_{i,j}^{II} \cdot (p_{idle}^{II})^j\right\} \notag \\
	&\quad+ \sum_{k=1}^{m} \sum_{i''=k}^{m} \sum_{j=W_{i''-1}}^{W_{i''}-1} \sum_{i=k}^{m} b_{i'',0}^{I} \cdot (p_{idle}^{I})^j \notag \\ 
	&\quad\cdot \left\{ (1-\gamma) \cdot b_{i,j}^{I} (p_{idle}^{I})^j + \gamma \cdot b_{i,j}^{II} (p_{idle}^{II})^j \right\}.
\end{align}
%\textcolor{blue}{where $p_{\text{Idle}}^{I}$ and $p_{\text{Idle}}^{II}$ represent the probability that the channel is idle per slot for each case respectively, which are given as }
%\begin{align}
%	p_{\text{Idle}}^{I} = (1 - \tau_{\text{AP}}^{\text{SLD}})^N \cdot (1 - \tau_{\text{MLD}}^I)^N 
%	\cdot (1 - \tau_{\text{SLD}}^I)^N,
%\end{align}
%\begin{align}
%	p_{\text{Idle}}^{II} = (1 - \tau_{\text{AP}}^{\text{MLD}})^N \cdot (1 - \tau_{\text{MLD}}^I)^N \cdot (1 - \tau_{\text{SLD}}^I)^N.
%\end{align}
%
In addition, when Link 1 is in Case 2, $Y_{{case}_2}$ is given as
\begin{align}
Y_{{case}_2} &= \sum_{i''=0}^{m} \sum_{j=1}^{W_0 -1} \sum_{i=0}^{m} b_{i'',0}^{II} \cdot (p_{idle}^{II})^j \notag \\ 
	&\quad\cdot \left\{ (1-\gamma) \cdot b_{i,j}^{I} \cdot (p_{idle}^{I})^j + \gamma \cdot b_{i,j}^{II} \cdot (p_{idle}^{II})^j\right\} \notag \\
	&\quad+ \sum_{k=1}^{m} \sum_{i''=k}^{m} \sum_{j=W_{i''-1}}^{W_{i''}-1} \sum_{i=k}^{m} b_{i'',0}^{II} \cdot (p_{idle}^{II})^j \notag \\ 
	&\quad\cdot \left\{ (1-\gamma) \cdot b_{i,j}^{I} (p_{idle}^{I})^j + \gamma \cdot b_{i,j}^{II} (p_{idle}^{II})^j \right\}.
\end{align}

%-------------------------------------------------------
\subsection{Derivation of per-device per-link throughput}

Finally, we derive per-device per-link throughput as follows. 
{\bf Per-link UL throughput of a legacy device:}
\begin{align}
\label{eq:legacy_UL}
S_U^{\text{SLD}} &= \frac{1}{N_{\text{SLD}}} \cdot \left[ (1-\gamma)\cdot\frac{\tau_{2}^{I}}{\phi_{case_1}}+\gamma\cdot\frac{\tau_{2}^{II}}{\phi_{case2}} \right] \cdot n_a \cdot L_D ,
\end{align}
{\bf Per-link UL throughput of a non-AP MLD:}
\begin{align}
\label{eq:MLD_UL}
S_U^{\text{MLD}} &= \frac{1}{N_{\text{MLD}}} \cdot \left[ (1-\gamma) \cdot \frac{\tau_{3}^{I}}{\phi_{case_1}} \cdot (1-p_{\text{MLD}}^I)^{N_{th}(p_{\text{MLD}}^I)} \right. \notag\\ 
	&\quad+ \left. \gamma \cdot \frac{\tau_{3}^{II}}{\phi_{case_2}} \cdot (1-p_{\text{MLD}}^{II})^{N_{th}(p_{\text{MLD}}^{II})} \right] 
	\cdot n_a\cdot L_D ,
\end{align}
{\bf Per-link DL throughput from AP MLD to a legacy device:}
\begin{align}
\label{eq:legacy_DL}
S_D^{\text{SLD}} =\frac{\tau_{1}^{I}}{\phi_{case_1}}\cdot n_a\cdot L_D ,
\end{align}
{\bf Per-link DL throughput from AP MLD to a non-AP MLD:}
\begin{align}
\label{eq:MLD_DL}
S_D^{\text{MLD}} &= \bigg[ \frac{\tau_{1a}^{II}}{\phi_{case_1}} + \frac{\tau_{1b}^{II}}{\phi_{case_2}} \cdot \left\{1-\frac{N_{th}(p_{\text{AP}}^{\text{MLD}})\cdot T_{empty}}{T_{data}}\right\}  \notag \\
	&\quad \cdot (1-p_{\text{AP}}^{\text{MLD}})^{N_{th}(p_{\text{AP}}^{\text{MLD}})} \Big] \cdot n_a\cdot L_D ,
\end{align}
where $n_a$ is the number of aggregated MPDUs, $L_D$ is the payload length, and $N_{th}(\cdot)$ is the average difference in backoff counter values between two links belonging to an MLD \cite{song2021performance}.

%% file: evaluation.tex
\begin{figure}
%\centerline{\includegraphics[width=0.8\linewidth]{figures/Fig9_throughput_same_num.eps}}
\centerline{\includegraphics[width=0.8\linewidth]{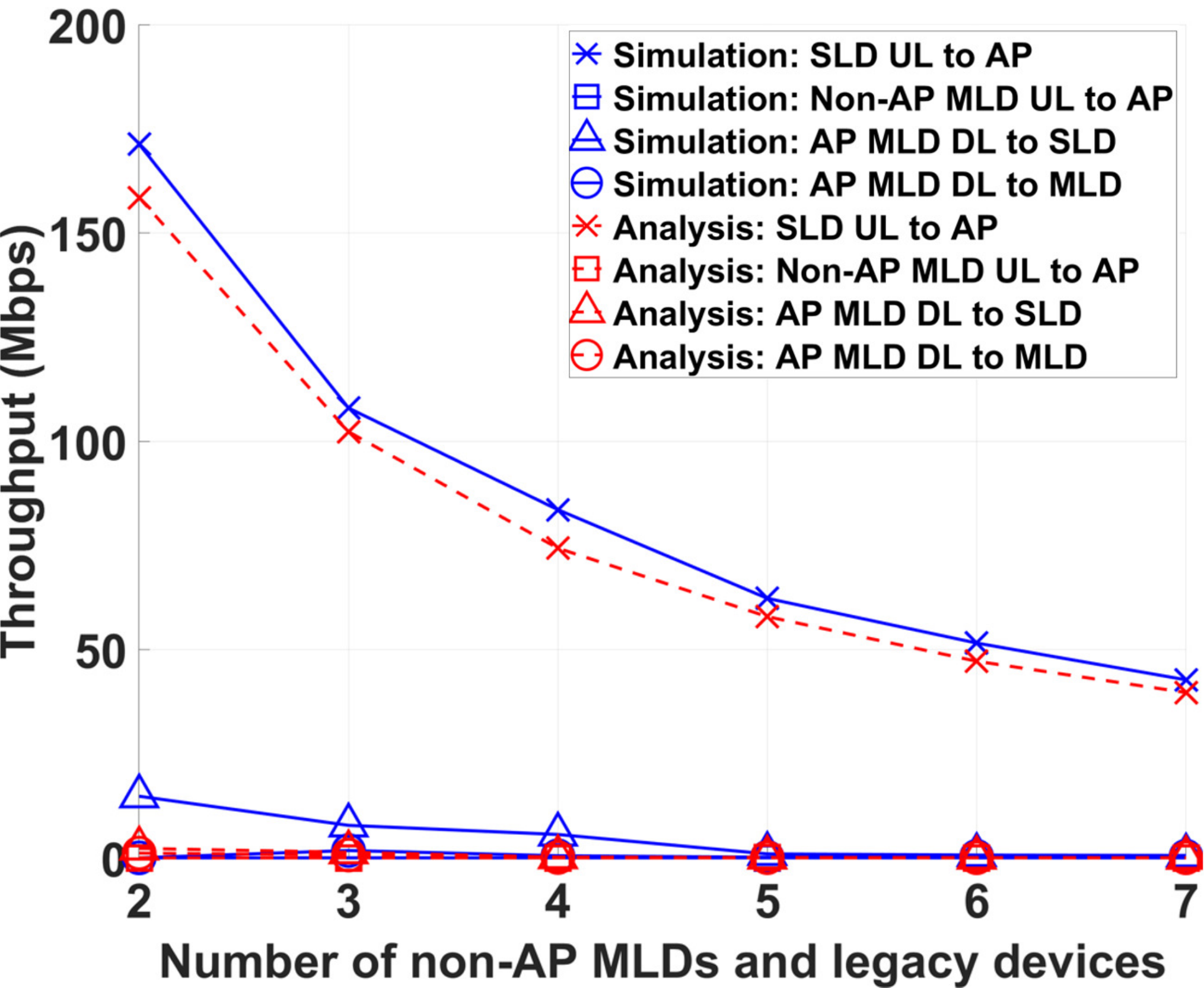}}
	\caption{Per-device throughput when $N_{\text{MLD}} = N_{\text{SLD}}$}
	\label{fig:throughput_same_num}
	\vspace{-0.1in}
\end{figure}

\begin{figure}
%\centerline{\includegraphics[width=0.8\linewidth]{figures/Fig10_throughput_diff_num.eps}}
\centerline{\includegraphics[width=0.8\linewidth]{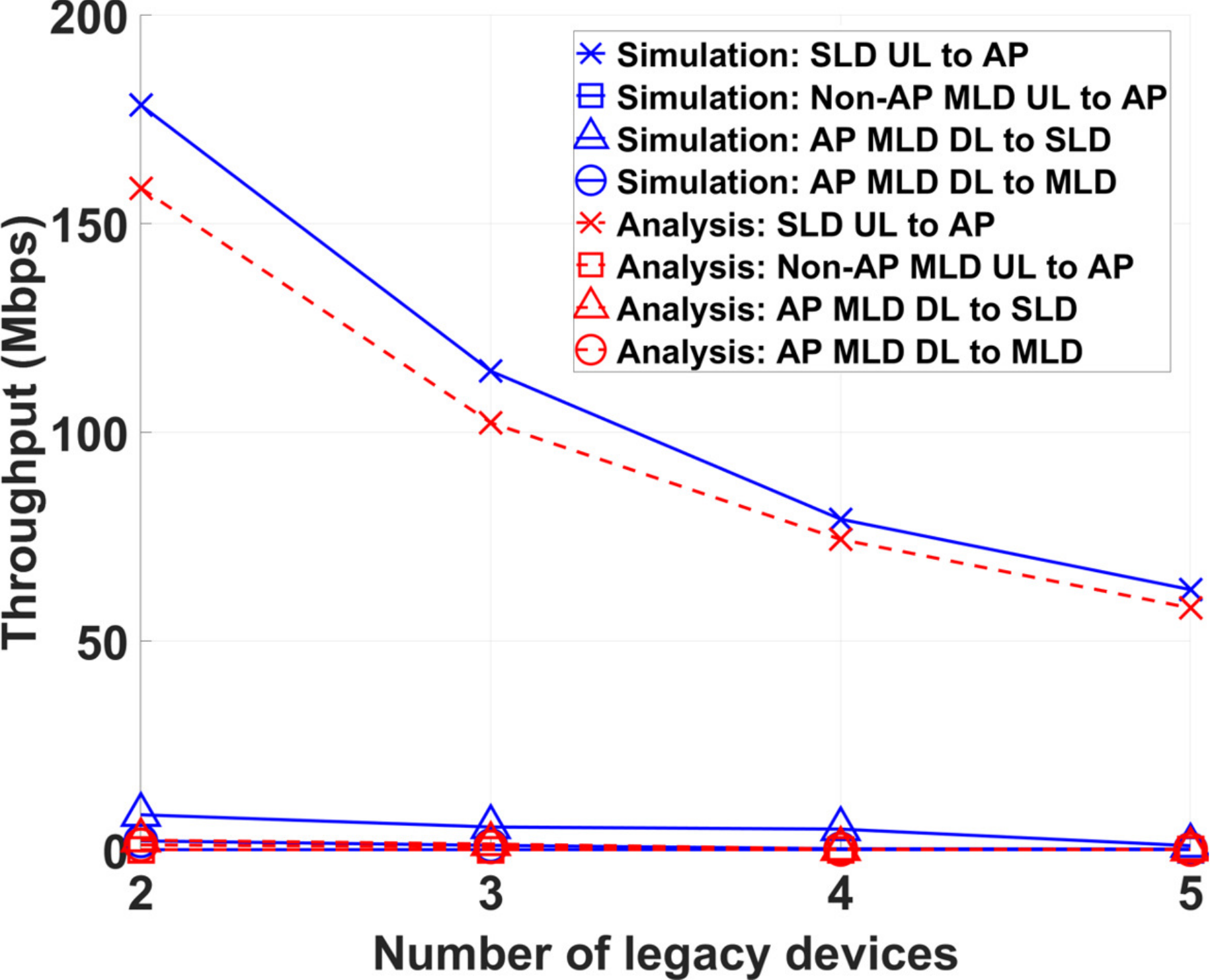}}
	%\caption{\textcolor{blue}{Throughput Analysis and Simulation Comparison ($N_{\text{MLD}} \neq N_{\text{SLD}}$)}}
	\caption{Per-device throughput when  $N_{\text{MLD}} \neq N_{\text{SLD}}$ and  $N_{\text{MLD}} = 5$}
	\label{fig:throughput_diff_num}
	\vspace{-0.1in}
\end{figure}

\begin{figure}
%\centerline{\includegraphics[width=0.8\linewidth]{figures/Fig11_throughput_diff_num_7.eps}}
\centerline{\includegraphics[width=0.8\linewidth]{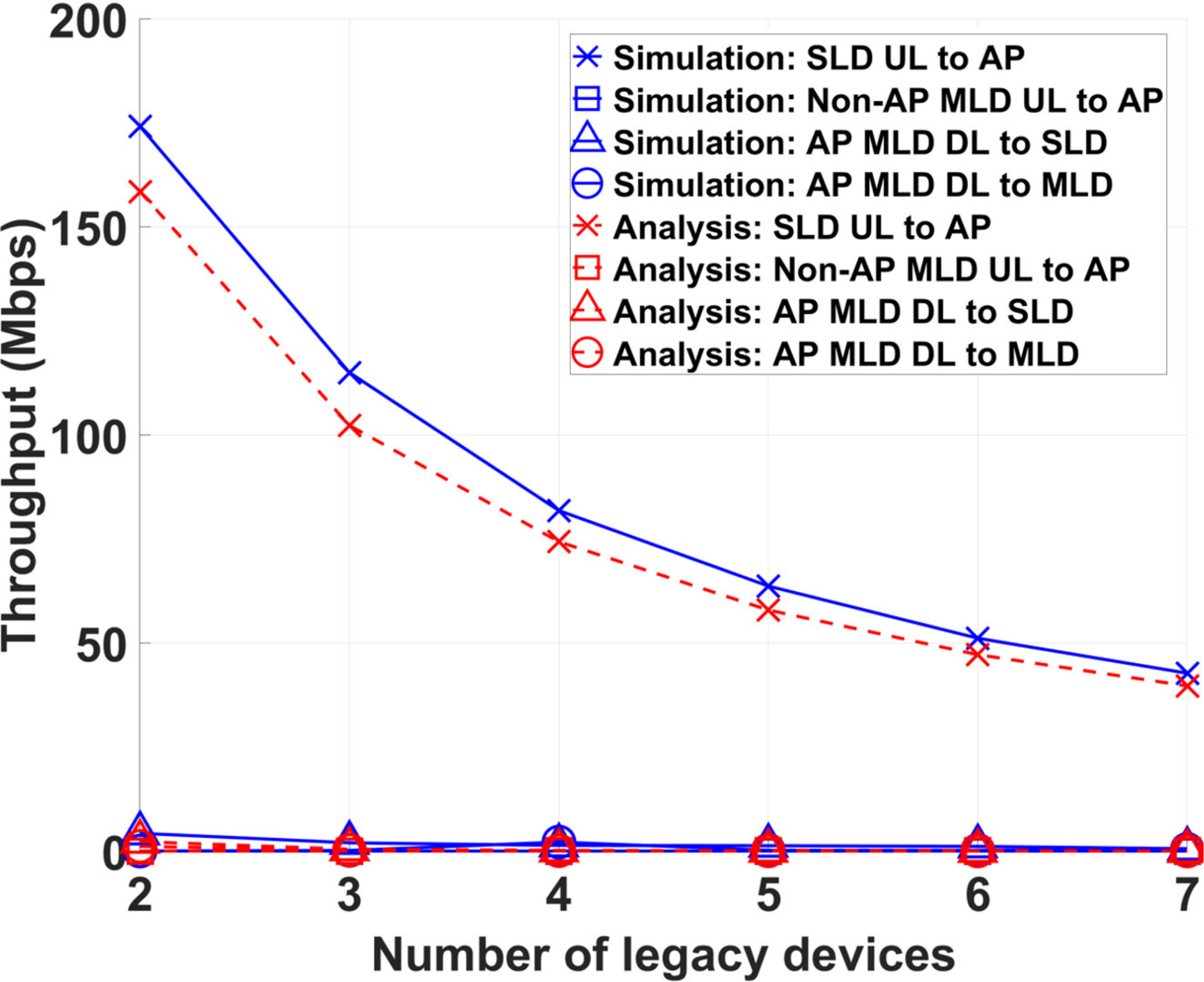}}
	%\caption{\textcolor{blue}{Throughput Analysis and Simulation Comparison ($N_{\text{MLD}} \neq N_{\text{SLD}}$)}}
	\caption{Per-device throughput when  $N_{\text{MLD}} \neq N_{\text{SLD}}$ and $N_{\text{MLD}} = 7$}
	\label{fig:throughput_diff_num_7}
	\vspace{-0.1in}
\end{figure}

This section evaluates the accuracy of the proposed analytical framework, and also examines the behavior of MLDs and legacy devices in various coexistence scenarios. %using numerical simulations.
For this purpose, we have jointly utilized two discrete-event network simulators, ns-3 \cite{henderson2008network} and MATLAB \cite{MATLAB}, where ns-3 is a modular C++ based simulator widely used in the networking field. %enabling detailed modeling of realistic network environments along with various performance metrics, making it a valuable tool for network research and analysis.
As of its latest release (ns-3.42), however, only STR is implemented among MLO functionalities, and hence we implemented essential NSTR functionalities in ns-3 to validate our proposed analysis, including start-time alignment, end-time alignment, and other key features proposed in this paper.
%
%Specifically, for the AP MLD, when transmitting packets to a non-AP MLD, the backoff process is reinitiated if an IDC interference issue is anticipated. 
%Additionally, when both links transmit packets to the same non-AP MLD, end-time alignment is enforced to ensure coordinated transmission completion.
%For non-AP MLDs, we implement start-time alignment, allowing transmissions to occur only when the backoff counters of both links reach zero simultaneously. 
%Furthermore, end-time alignment is incorporated.
%
In addition, to evaluate $\tau$'s and $p$'s which are iteratively affecting each other as shown in Sections \ref{sec:ap_mc} through \ref{sec:analysis}, 
%As observed in the derivation process from Section \ref{sec:ap_mc} to Section \ref{sec:analysis}, $\tau$'s and $p$'s influence each other iteratively.
%To calculate $\tau$'s and $p$'s in a given network environment and use them to determine throughput, 
we utilized MATLAB's Optimization Toolbox and its nonlinear equation solver \cite{OptimizationToolbox}. 
%These approaches allow us to compare the throughput performance of coexisting AP MLD, non-AP MLDs, and legacy devices, providing an in-depth analysis of their behavior and performance in coexisting network environments.

As described earlier in Section \ref{sec:model}, we assume that each MLD has two links at 5 GHz and 6 GHz, respectively, and the network consists of one AP MLD, $N_{\text{MLD}}$ non-AP MLDs, and $N_\text{legacy}$ legacy devices per link.
All devices are randomly distributed within the 5-meter radius from the AP and use MCS 8 with 80 MHz channel bandwidth.
We also assume the saturated traffic condition, and hence all devices always have packets to transmit where the packet size is 1500 bytes.
%\textcolor{red}{To ensure no packet expiration in the MAC queue, the packet lifetime is set to a sufficiently large value of 10,000 ms \cite{magrin2023performance}.} 
In addition, each simulation scenario was run for 5 more seconds after every device's queue is sufficiently filled with packets. 

\textcolor{black}{Note that while our analytical model focuses on dual-link MLDs, it does not assume symmetric link behaviors at all.
Unlike \cite{song2021performance}, we construct separate per-link MCs  to model independent contention/channel dynamics.
Such flexibility for heterogeneity has also been reflected in our simulation setup.}

%-------------------------------------------------------
\subsection{Per-device throughput when $N_{\text{MLD}} = N_{\text{SLD}}$}
\label{subsec:same_num}

First, we compare per-device per-link throughput  obtained from our analysis with the one measured by ns-3 simulation, for scenarios with $N_{\text{MLD}} = N_{\text{SLD}}$.
Fig. \ref{fig:throughput_same_num} illustrates the throughput trend as $N_{\text{MLD}}$ and $N_{\text{SLD}}$ increase up to seven.
The plots show that 
all three types of throughput (i.e., SLD's, Non-AP MLD's, and AP MLD's) tend to decrease with $N_{\text{MLD}}$ (or equivalently $N_{\text{SLD}}$) due to aggravated competition among the coexisting devices, while
the proposed analysis and simulated results closely correspond to each other for each type of devices.
Among them, the SLD's (i.e., legacy device's) UL throughput achieves a dominantly large throughput, where the discrepancy between analysis and simulation presents an acceptable level of error varying as $[7.4, 5.2, 11.0, 7.0, 8.4, 7.1]$ (\%) for $N_{\text{MLD}} = N_{\text{SLD}} = [2, 3, 4, 5, 6, 7]$ respectively.
The AP MLD's DL throughput and the non-AP MLD's UL throughput, however, are evaluated as significantly smaller than the SLD UL throughput, making the comparison between analysis and simulation less meaningful.
Nevertheless, all the results confirm that the gap between analysis and simulation gets smaller as the number of non-AP STAs increases, showing the reliability of the proposed analysis even in a larger network.

More specifically, the SLD UL throughput achieves $[158.4, 102.2, 74.3, 57.8, 47.1, 39.6]$ Mbps in analysis and $[171.2, 107.9, 83.5, 62.2, 51.5, 42.7]$ Mbps in simulation.
In the meantime, the non-AP MLD UL throughput remains negligible, ranging from $0.000007$ to $0.00247$ Mbps across all cases in analysis, %[0.0027, 0.00002, 0.00002, 0.00001, 0.0001, 0.000007, 0.000007]
	while it is consistently measured as $0$ Mbps in simulation.
For the AP MLD, its DL throughput to an SLD steadily decreases as $[2.2, 1.3, 0.1, 0.08, 0.04, 0.03]$ Mbps in analysis, while the simulation shows $[14.7, 7.7, 5.6, 0.9, 0.6, 0.5]$ Mbps. 
In addition, the AP MLD's DL throughput to a non-AP MLD presents $[1.1, 0.6, 0.08, 0.04, 0.02, 0.01]$ Mbps in analysis, while showing $[0, 1.7, 0.4, 0.1, 0.3, 0.3]$ Mbps in simulation.
%} 

%In the ns-3 simulation, the legacy devices achieve slightly higher uplink throughput compared to the analysis, measured at $[171.27, 107.95, 83.55, 62.28, 51.55, 42.72]$ Mbps.
%Non-AP MLDs, however, fail to secure any transmission opportunities, resulting in a consistent throughput of 0 Mbps. 
%For AP MLD's downlink throughput to a legacy device, the simulation results are higher than the analysis, with values of $[14.78, 7.79, 5.6, 0.94, 0.66, 0.56]$ Mbps.
%The throughput for transmission to non-AP MLDs is extremely limited, with values between 0 Mbps and 1.75 Mbps, represented by $[0, 1.75, 0.49, 0.14, 0.36, 0.39]$ Mbps.

The aforementioned results reveal significant challenges in utilizing MLO in the coexistence scenario with Wi-Fi 7 and legacy Wi-Fi.
Employing non-AP MLDs becomes disadvantageous particularly due to its start-time alignment requirement, which mandates that both links have their backoff counters reach zero simultaneously. 
In saturated conditions, however, all the channels are frequently busy, causing one link to restart its backoff process when the other link switches from idle to busy. %encounters channel activity.
As a result, legacy devices may dominate to utilize the channels securing the majority of transmission opportunities, while MLDs are struggling to obtain the chance for transmission.
On the other hand, the AP MLD also faces difficulty in transmitting packets to non-AP MLDs.
If the AP attempts transmission to a non-AP MLD on one link, it should not proceed if the other link is busy except when its busyness is due to the transmission by the AP itself, so as to prevent potential IDC interference at non-AP MLDs as discussed  in Section \ref{sec:model}.
Moreover, saturated conditions exacerbate such a situation, causing the packets destined to non-AP MLDs to remain in the AP MLD's queue, ultimately leading to a reduced DL throughput. 

%The dominance of legacy devices in securing transmission opportunities explains their higher throughput. 
%The discrepancy between analysis and simulation results for legacy device throughput varies across scenarios, with error percentages of $[7.48, 5.27, 11, 7.05, 8.48, 7.14]$ (\%) for $N_{\text{MLD}} = N_{\text{SLD}} = [2, 3, 4, 5, 6, 7]$, respectively.
%The highest error occurs at 11\% when $N_\text{MLD} = N_\text{SLD} = 4$, while in all other cases, the error remains below 8.5\%, demonstrating overall consistency between the analysis and simulation results.
%Notably, the throughput of the AP MLD and non-AP MLDs is significantly smaller compared to legacy devices, making error rate comparisons for these devices less meaningful.
%These results suggest that the analytical model and simulations provide reasonably consistent outcomes for scenarios involving larger numbers of devices.

%-------------------------------------------------------
\subsection{Per-device throughput when $N_{\text{MLD}} \neq N_{\text{SLD}}$}

%We observed in Section \ref{subsec:same_num} that when the number of non-AP MLDs and legacy devices is equal, legacy devices dominate transmissions, leaving little to no transmission opportunities for the AP MLD and non-AP MLDs.
To investigate how the system dynamics change when there exist an imbalanced number of MLDs and SLDs, i.e., $N_{\text{MLD}} \neq N_{\text{SLD}}$, we now fix $N_{\text{MLD}}$ to 5, and vary the number of legacy devices from 2 to 5.
Likewise, we also consider the case with $N_{\text{MLD}} = 7$ and $N_{\text{SLD}}$ varies from 2 to 7. 
%, and the case with $N_{\text{MLD}} = 7$ and $N_{\text{SLD}}$ varies from two to seven. 
%
%Using such setups, we calculate and observe the throughput using both the analytical model and ns-3 simulations.

Fig. \ref{fig:throughput_diff_num} shows that even when $N_{\text{SLD}}$ is fewer than $N_{\text{MLD}}$, the results remain similar to those observed in Fig. \ref{fig:throughput_same_num}. 
More specifically, the SLD UL throughput achieves $[158.4, 102.2, 74.3, 57.8]$ Mbps in analysis and $[178.3, 114.6, 79.1, 62.2]$ Mbps in simulation.
In the meantime, the non-AP MLD UL throughput remains negligible, yielding values that are effectively zero in analysis and exactly zero in simulation.
For the AP MLD, its DL throughput to an SLD steadily decreases as $[2.2, 1.3, 0.1, 0.08]$ Mbps in analysis, while the simulation shows $[8.3, 5.4, 4.9, 0.9]$ Mbps. 
In addition, the AP MLD's DL throughput to a non-AP MLD presents $[1.1, 0.6, 0.06, 0.04]$ Mbps in analysis, while showing $[2.0, 1.0, 0.2, 0.1]$ Mbps in simulation.
%
%\textcolor{blue}{Fig. \ref{fig:throughput_diff_num} shows that even when the number of legacy devices is fewer than the number of non-AP MLDs, the results remain similar to those observed in Fig. \ref{fig:throughput_same_num}. 
%According to the analysis, the per-device throughput for legacy devices is calculated as $[158.46, 102.26, 74.31, 57.89]$ Mbps. 
%In contrast, the uplink throughput for non-AP MLDs remains negligible across all cases, yielding values that are effectively zero.
%For the AP MLD, the downlink throughput to a legacy device is computed as $[2.25, 1.34, 0.167, 0.082]$ Mbps, while the throughput to a non-AP MLD is $[1.13, 0.67, 0.067, 0.041]$ Mbps.
%}
%\textcolor{blue}{In the ns-3 simulation, the throughput for legacy devices is slightly higher than the analytical results, measured at $[178.33, 114.62, 79.16, 62.28]$ Mbps. 
%Non-AP MLDs, however, consistently achieve an uplink throughput of zero.
%The AP MLD's downlink throughput to a legacy device is recorded as $[8.36, 5.42, 4.92, 0.94]$ Mbps, while the throughput to a non-AP MLD is $[2.09, 1.02, 0.23, 0.14]$ Mbps.
%}

A notable observation here is that even when legacy devices are less populated than non-AP MLDs, the latter still hardly transmit.
This result demonstrates that under saturated conditions, satisfying the start-time alignment requirement becomes really challenging.
The same observation can be made for the AP MLD as discussed in Section \ref{subsec:same_num}, which faces the challenge in avoiding potential IDC interference. 
%Similarly, as observed in Section \ref{subsec:same_num}, the AP MLD faces significant challenges in transmitting effectively because of IDC interference issues. 
%These issues prevent the AP MLD from transmitting smoothly when packets destined for MLDs are queued.
%
In the meantime, the error between analysis and simulation for the SLD UL throughput is measured as $[11.1, 10.7, 6.1, 7.0]$ (\%) for $N_{\text{SLD}} = [2, 3, 4, 5]$ respectively, demonstrating a reasonable level of consistency.

%To further analyze how system performance changes with different ratios of non-AP MLDs to legacy devices, we now consider the case where the number of non-AP MLDs is fixed at six, and the number of legacy devices varies from two to six. 
Next, Fig. \ref{fig:throughput_diff_num_7} illustrates the case when $N_{\text{MLD}} = 7$ and $N_{\text{SLD}}$ varies from 2 to 7, which allows us to explore whether increasing the number of non-AP MLDs alters the dynamics observed in previous setup.
%the per-device throughput dynamics for each device type in the scenario where $N_{\text{MLD}}$ is fixed at six and $N_{\text{SLD}}$ varies from two to six.
%
Similar to the previous scenario, the SLD UL throughput dominates the network while the AP MLD DL throughput is significantly smaller and the non-AP MLD UL throughput stays negligible.
Particularly, the SLD UL throughput is obtained as $[158.4, 102.2, 74.3, 57.9, 47.1, 39.6]$ Mbps by analysis while the simulation measures slightly higher values of $[174.1, 114.9, 81.8, 63.7, 51.1, 42.7]$ Mbps, presenting the error of $[9.0, 11.1, 9.2, 9.1, 7.8, 7.3]$ (\%), still demonstrating a reasonable level of consistency. 
Such results indicate that under saturated conditions, despite of the further increase in the number of non-AP MLDs, the presence of even a small number of legacy devices prevents both the AP MLD and non-AP MLDs from obtaining meaningful transmission opportunities due to the constraints imposed by NSTR operations.

\begin{comment}
%
For a legacy device, the throughput calculated by the analysis is $[158.46, 102.26, 74.31, 57.90, 47.18]$ Mbps, while the ns-3 simulation measures slightly higher values of $[178.07, 115.34, 80.48, 63.04, 51.55]$ Mbps.
The error rate between the analytical and simulation results for legacy device throughput are $[11.01, 11.34, 7.67, 8.15, 8.48]$ (\%), demonstrating a reasonable level of consistency. 
As observed in previous analyses, the throughput of AP MLD remains significantly smaller than that of a legacy device in both the analysis and simulation. 
Similarly, the throughput of non-AP MLDs is negligible in the analysis and consistently observed as zero in all cases in the ns-3 simulations. 
Due to the extremely low throughput of the AP MLD and a non-AP MLD, which aligns with previous results, detailed numerical values for MLDs are omitted.

These results indicate that even when the number of non-AP MLDs increases from five to six, the presence of even a small number of legacy devices under saturated conditions prevents both the AP MLD and non-AP MLDs from obtaining meaningful transmission opportunities due to the constraints imposed by NSTR operations.
\end{comment}

%% file: conclusion.tex
This paper addressed some critical operations of MLO overlooked by existing studies, specifically focusing on the backoff behavior of MLDs as mandated by the standard.
%This paper delved into the challenges of implementing MLO in Wi-Fi 7 devices, exploring essential considerations for channel access.
%Our investigation extended to overlooked aspects in existing studies, specifically focusing on the backoff behavior of MLDs as mandated by standards.
In particular, we proposed a series of novel MCs that accurately model the behavior of the standard-compliant AP MLD and non-AP MLDs respectively, and derived their stationary probabilities and closed-form per-device throughputs in the Wi-Fi 7 coexistence scenario.
%Building upon this analysis, we aim to finalize the development of Markov Chains for each type and progress to crafting algorithms to enhance fairness in throughput within coexistence scenarios.

In future, we would like to leverage on the developed framework to develop more efficient coexistence mechanisms for MLO-based Wi-Fi 7 with legacy WLANs.

%% file: bios.tex
%\newpage
%\vspace{0.1in}

% BIOGRAPHY - S Jung
\begin{IEEEbiography}
%[{\includegraphics[width=1in,clip,keepaspectratio]{figures/Fig_SJung.eps}}]
[{\includegraphics[width=1in,clip,keepaspectratio]{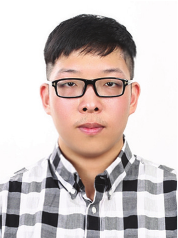}}]
{Suhwan Jung} is a Ph.D. candidate at the Department of Electrical Engineering (EE), the Ulsan National Institute of Science and Technology (UNIST), Ulsan, Korea.
He received his B.S. degree in the School of Electrical and Computer Engineering (ECE) from the UNIST in 2017, and his M.S. degree in the Department of Electrical Engineering (EE) from the UNIST in 2019.
His research interest includes 5G/6G, V2X, and WLAN Multi-Link Operation (MLO).
\end{IEEEbiography}

% BIOGRAPHY - S Choi
\begin{IEEEbiography}
[{\includegraphics[width=1in,clip,keepaspectratio]{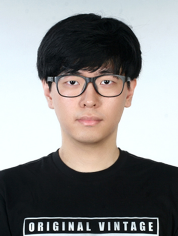}}]
{Seokwoo Choi} is a postdoctoral researcher at the School of Electrical Engineering, the Ulsan National Institute of Science and Technology (UNIST), Ulsan, Korea, since March 2025.
He received his B.S. degree in Electrical and Computer Engineering from the UNIST, in 2017, and received Ph.D. degree in Electrical Engineering from UNIST, in 2025.
His research interest includes 6G communications, WLAN, AI/ML, and low latency networking.
\end{IEEEbiography}

% BIOGRAPHY - Y Yoon
\begin{IEEEbiography}[{\includegraphics[width=1in,clip,keepaspectratio]{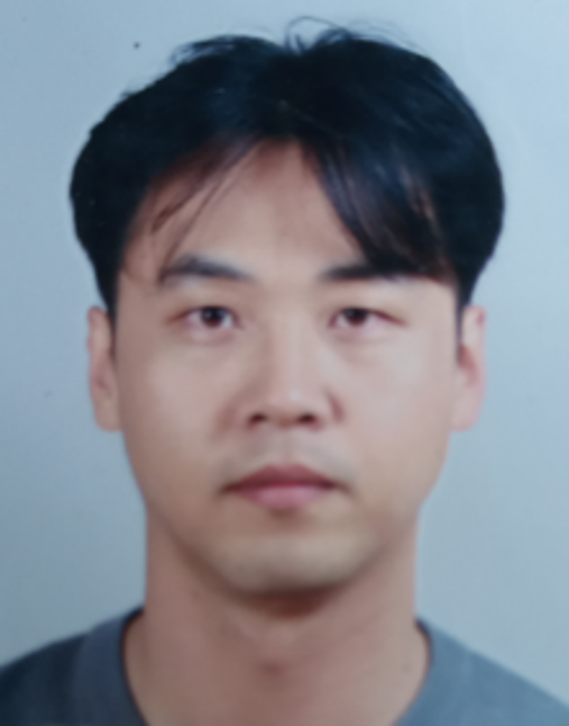}}]
{Youngkeun Yoon} received the B.E. and M.E. degrees in radio engineering from Chungbuk National University, Cheongju, South Korea in 1997, 1999, respectively. 
Since 2000, he has been worked in Electronics and Telecommunications Research Institute (ETRI). 
He has been involved in the research of simulation and evaluation for the radio resource management based on the multiple cellular system since 2003. 
He also was a visitor researcher in Center for Communication Systems Research (CCSR), University of Surrey, UK from 2003 to 2004. 
He has been involved in the analysis of the impacts for interference between existing system and new system. 
Currently, his research interests are in system performance and evaluations, radio propagation, and artificial intelligence based on radio communication system.
\end{IEEEbiography}

% BIOGRAPHY - Hk Son
\begin{IEEEbiography}
[{\includegraphics[width=1in,clip,keepaspectratio]{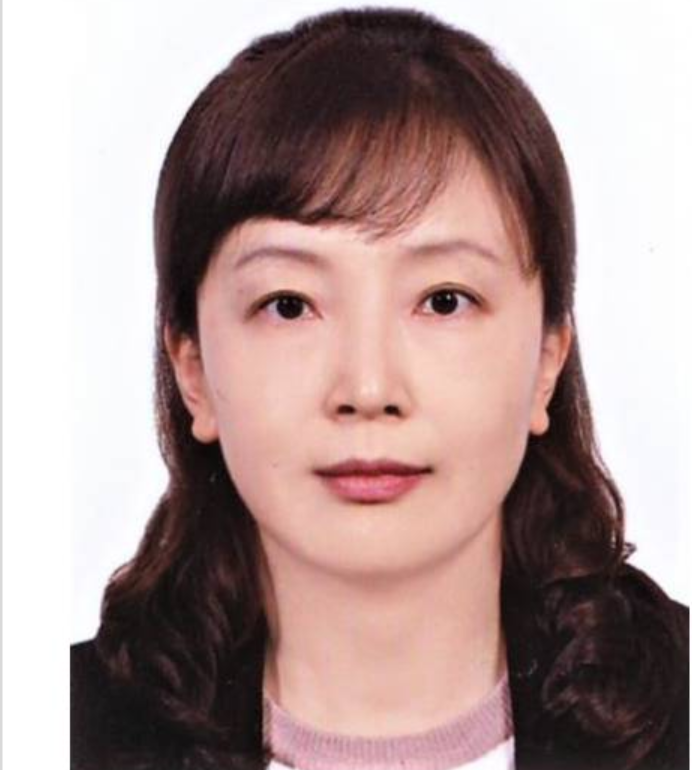}}]
{Ho-kyung Son} received the B.S and M.S degrees in  Electronic Engineering from Kyungpook National  University, Daegu, Korea, in 1997 and 1999, and the  Ph.D degree in the same university, Korea in 2013. 
From March 1999 to March 2000, she was a  researcher of LG Information and Communication.  
Since May 2000, she has been with the Electronics  and Telecommunications Research Institute (ETRI),  Daejeon, Korea, where she is a principal researcher of radio resource research section of the Radio Research Division. 
Her research interests include radio resource management, interference analysis between wireless communications, and radio  propagation.
\end{IEEEbiography}

% BIOGRAPHY - H Kim
\begin{IEEEbiography}[{\includegraphics[width=1in,clip,keepaspectratio]{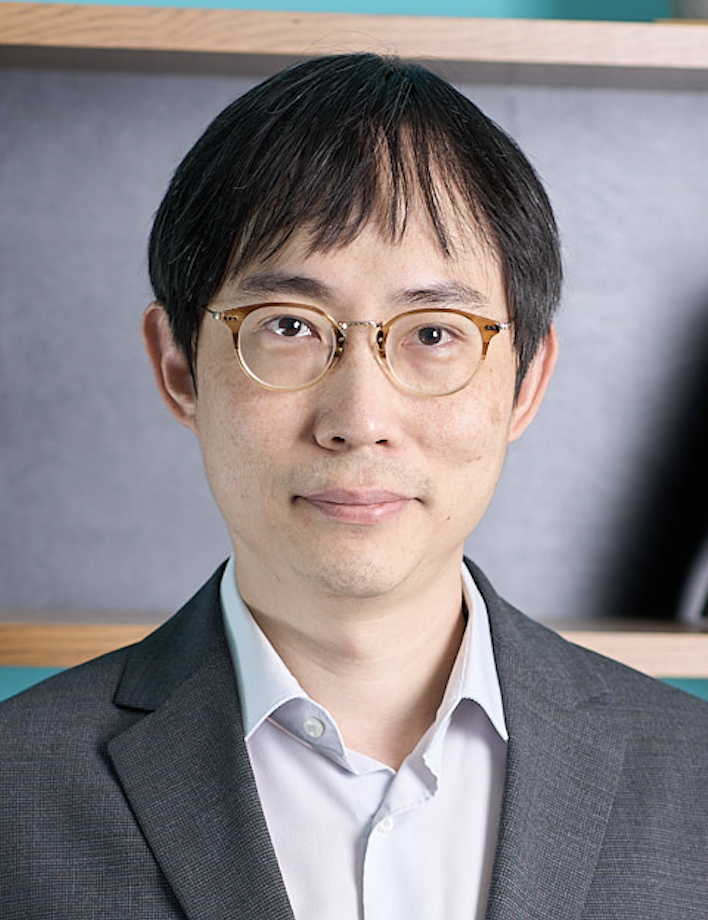}}]
{Hyoil Kim} is a professor at the Department of Electrical Engineering, Ulsan National Institute of Science and Technology (UNIST), Korea.
He received his B.S. degree in Electrical Engineering from Seoul National University in 1999, and M.S. and Ph.D. degrees in Electrical Engineering: Systems from the University of Michigan in 2005 and 2010, respectively.
Before joining UNIST in 2011, he was a postdoctoral researcher in the IBM T.J. Watson Research Center, Hawthorne, NY, USA in 2010--2011.
His research interest includes Hyperloop communications, B5G/6G, O-RAN, WLAN, AI/ML, V2X, UAM/satellite communications, and Dynamic Spectrum Access.
%He has served as a TPC member for various international conferences, including IEEE INFOCOM (2014, 2016-2017, 2021-2022), IEEE SECON (2020-2021), IEEE Globecom (2011-2015, 2017-2018). 
%He is an Area Editor of Computer Networks (ComNet) and an Editor of the Journal of Communications and Networks (JCN). 
%He also served as a Guest Editor for Ad Hoc Networks and the IEEE Trans. on Cognitive Communications and Networking (TCCN).
\end{IEEEbiography}